%% file: neurips_2024.tex
\documentclass{article}

    \PassOptionsToPackage{numbers, compress}{natbib}

\usepackage[final]{neurips_2024}

\usepackage[utf8]{inputenc} 
\usepackage[T1]{fontenc}    
\usepackage{hyperref}       
\usepackage{url}            
\usepackage{booktabs}       
\usepackage{amsfonts}       
\usepackage{nicefrac}       
\usepackage{microtype}      
\usepackage{xcolor}         
\usepackage{graphicx}
\graphicspath{ {./figures/} }
\usepackage{caption}
\usepackage{array}
\usepackage{amsmath}
\usepackage{amsthm}
\usepackage[ruled]{algorithm2e}
\usepackage[normalem]{ulem}

\SetKwComment{Comment}{/* }{ */}

\newcommand{\PreserveBackslash}[1]{\let\temp=\\#1\let\\=\temp}
\newcolumntype{C}[1]{>{\PreserveBackslash\centering}p{#1}}
\newcolumntype{R}[1]{>{\PreserveBackslash\raggedleft}p{#1}}
\newcolumntype{L}[1]{>{\PreserveBackslash\raggedright}p{#1}}
\include{commands}

\title{RandNet-Parareal: a time-parallel PDE solver using Random Neural Networks}

\author{%
Guglielmo Gattiglio\\
Department of Statistics\\
University of Warwick\\
Coventry, CV4 7AL, UK\\
\texttt{Guglielmo.Gattiglio@warwick.ac.uk}\\
\And
  Lyudmila Grigoryeva\thanks{Honorary Associate Professor, Department of Statistics, University of Warwick, Coventry, CV4 7AL, UK. (\texttt{Lyudmila.Grigoryeva@warwick.ac.uk})} \\
  Faculty of Mathematics and Statistics \\
  University of St.~Gallen \\
  Rosenbergstrasse~20, CH-9000 St.~Gallen,~Switzerland\\
  \texttt{Lyudmila.Grigoryeva@unisg.ch} \\
  \And
  Massimiliano Tamborrino \\
  Department of Statistics\\
University of Warwick\\
Coventry, CV4 7AL, UK\\
  \texttt{Massimiliano.Tamborrino@warwick.ac.uk} \\
}

\begin{document}

\maketitle
\begin{abstract}
Parallel-in-time (PinT) techniques have been proposed to solve systems of time-dependent differential equations by parallelizing the temporal domain. Among them, Parareal computes the solution sequentially using an inaccurate (fast) solver, and then ``corrects'' it using an accurate (slow) integrator that runs in  parallel across temporal subintervals. This work introduces RandNet-Parareal, a novel method to learn the discrepancy between the coarse and fine solutions using random neural networks (RandNets). RandNet-Parareal achieves speed gains up to x125 and x22 compared to the fine solver run serially and Parareal, respectively. Beyond theoretical guarantees of RandNets as universal approximators, these models are quick to train, allowing the PinT solution of partial differential equations on a spatial mesh of up to $10^5$ points with minimal overhead, dramatically increasing the scalability of existing PinT approaches. RandNet-Parareal's numerical performance is illustrated on systems of real-world significance, such as the viscous Burgers' equation, the Diffusion-Reaction equation, the two- and three-dimensional Brusselator, and the shallow water equation.

\end{abstract}

\section{Introduction}
\label{sec:intro}
Parallel-in-time (PinT) methods %
have been used to overcome the saturation of  well-established spatial parallelism approaches for solving (prohibitively expensive)  initial value problems (IVPs) for ordinary %
and partial differential equations (ODEs and PDEs), described by 
systems of $d \in \mathbb{N}$ ODEs (and similarly for PDEs)
\begin{equation}
\label{eq:ode}
\frac{d \boldsymbol{u}}{d t}=h(\boldsymbol{u}(t), t) \enspace \text { on } t \in\left[t_0, t_N\right], \enspace \text {with } \boldsymbol{u}\left(t_0\right)=\boldsymbol{u}^0, \enspace N\in \mathbb{N},
\end{equation}
where $h: \mathbb{R}^d \times\left[t_0, t_N\right] \rightarrow \mathbb{R}^d$ is a smooth multivariate function, $\boldsymbol{u}:\left[t_0, t_N\right] \rightarrow \mathbb{R}^d$ is the time dependent column vector solution, and $\boldsymbol{u}^0 \in \mathbb{R}^d$ is the initial value at $t_0$. PinT schemes are particularly important when the sequential application of an accurate numerical integrator $\f$ over $\left[t_0, t_N\right]$ is infeasible in  a reasonable wallclock time. There are three general approaches for PinT computation: parallel across-the-problem, parallel-across-the-step, and parallel-across-the-method. In \cite{Gander2015,ong2020applications}, another classification is provided: multiple shooting, methods based on waveform relaxation and domain decomposition, multigrid approaches, and direct time-parallel methods. Parallel-across-the-step methods, in which solutions at multiple time-grid points are computed simultaneously, include Parareal (approximation of the derivative in the shooting method) \cite{lions2001resolution}, Parallel Full Approximation Scheme in Space and Time (PFASST) (multigrid method) \cite{emmett2012toward, minion2011hybrid}, and Multigrid Reduction in Time (MGRIT) \cite{Falgout2014, Friedhoff2012} methods (see \cite{Gander2023} for details). Among them, Parareal~\cite{lions2001resolution} has garnered popularity, with extensive theoretical analyses, improved versions, and empirical applications \cite{Gander2015, ong2020applications}. This is due to its non-intrusive nature which allows seamless integration with arbitrary temporal and spatial discretizations, and %
to its successful performance across diverse fields, 
such as plasma physics~\cite{ reynolds2012mechanisms,samaddar2010parallelization,samaddar2019application}, finance~\cite{bal2002parareal,pages2018parareal}, and weather modeling~\cite{philippi2022parareal,philippi2023micro}. 
Limited theoretical results are available for MGRIT and PFASST, with a few extensions and empirical applications. Interestingly, combined analyses have shown equivalences between Parareal and MGRIT, and connections between MGRIT and PFASST.
In Parareal, a coarse and fast solver $\g$ is run sequentially to obtain a first approximation of the solution, which is then corrected by running a fine (accurate) but slow integrator $\f$ in parallel across $N$ temporal subintervals. This procedure is then iterated until a convergence criterion is met after $k \leq N$ iterations, leading to a speed-up compared to running $\f$ sequentially over the entire time interval.
A recent advancement, GParareal~\cite{pentland2023gparareal}, improves Parareal convergence rates (measured as $k/N$) by learning the discrepancy $\f-\g$ using Gaussian Processes (GPs). This method outperforms Parareal for low-dimensional ODEs and a moderate number of computer cores $N$. However, the cubic cost (in the number of data points, roughly $kN$ at iteration $k$) of inverting the GP covariance matrix hinders its broader application. Subsequent research introduced nearest neighbors (nns) GParareal (nnGParareal)~\cite{nngparareal}, enhancing GParareal's scalability properties in both $N$ and $d$ through data reduction. Significant computational gains were achieved by training the GP on a small subset of nns, resulting in an algorithm loglinear in the sample size. 
This allowed scaling its effectiveness up to
systems 
with a few thousand ODEs, beyond which it loses its potential. Indeed, being based on the original GP framework, it uses a costly hyperparameter optimization procedure that requires fitting one GP per ODE dimension.

This study introduces RandNet-Parareal, a new approach using random neural networks (RandNets) to learn the discrepancy $\f-\g$. RandNets are a family of single-hidden-layer feed-forward neural networks (NNs), where hidden layer weights are randomly sampled and fixed, and only the output (or readout) layer is subject to training. Compared to standard artificial NNs, RandNets are hence much simpler to train: the input data are fed through the network, the predictions observed, and the weights of the linear output (or readout) layer are obtained as minimizers of a penalized squared loss between the NN outputs and the training targets. Since this optimization problem admits a closed-form solution, no backpropagation is required, and the issues of vanishing and exploding gradients persisting for standard fully trainable NNs are therefore avoided. The literature on the topic is rich and somewhat fragmented, and different names are used for essentially the same model. RandNets are  related to Random Feature Networks~\cite{Carratino2018,meimontanari2022,rahimi2008uniform,rahimi2009random,Rudi2016GeneralizationPO} and Reservoir Computing~\cite{RC8,RC24,gonon2023approximation,RC6,RC9}, Random Fourier Features (RFFs) and kernel methods~\cite{kar2012random,rahimi2007random,  sinha2016learning, sun2018but}. Some authors use the name Extreme Learning Machines (ELMs)~\cite{huang2014insight,huang2004extreme,huang2006universal,huang2006extreme, li2005fully} to refer to RandNets, while others use the term randomized or random NNs~\cite{cao2018review,he2016fuzzy,ji2015crucial,lu2014extended,  wan2015novel,  zhang2016survey} for the same paradigm. RandNets show excellent empirical performance, and have been used in the context of mathematical finance \cite{gonon2023BS,herrera2021,jacquier2023random}, mathematical  physics~\cite{neufeldschmocker2024}, electronic circuits~\cite{Caravelli2022}, photonic~\cite{lupo2021} and quantum systems~\cite{GononJacq2023,QRC1}, random deep splitting schemes~\cite{neufeld2024error}, scientific computing ~\cite{dong2021,dwivedi2020,dongwang2023,yang2018}, and have shown excellent empirical performance in numerous further applications. Moreover, recent work~\cite{gonon2023BS,gonon2023approximation} proves that RandNets are universal approximators within spaces of sufficiently regular functions, and provides explicit approximation error bounds, with these  results generalized to a large class of Bochner spaces in
\cite{neufeldschmocker2024}. 
These contributions show that  RandNets are a reliable machine learning paradigm with provable theoretical guarantees.

In this paper, we show that endowing Parareal with RandNets-based learning of  $\f-\g$,  the new proposed RandNet-Parareal algorithm, leads to significantly improved scalability, convergence speed, and parallel performance with respect to nnGParareal, GParareal, and Parareal. This allows us to solve PDE systems on a fine mesh of up to $10^5$ discretization points with negligible overhead, outperforming nnGParareal by two orders of magnitude and reducing its model cost by several orders.

Here, we compare the performance of Parareal, nnGParareal, and RandNet-Parareal on five increasingly complex systems, some of which are drawn from an extensive benchmark study of time-dependent PDEs~\cite{takamoto2022pdebench}. These include the one-dimensional %
viscous Burgers' equation, the two-dimensional %
Diffusion-Reaction equation, a challenging benchmark used to model biological pattern formation~\cite{Turing1952}, the two- and three-dimensional Brusselator, known for its complex behavior, including oscillations, spatial patterns, and chaos, and the shallow water equations (SWEs). Derived from the compressible Navier-Stokes equations, the SWEs are a system of hyperbolic PDEs exhibiting several types of real-world significance behaviors known to challenge numerical integrators, such as sharp shock formation dynamics, sensitive dependence on initial conditions, diverse boundary conditions, and spatial heterogeneity. Example applications include of tsunamis or flooding simulations.

We intentionally chose two hyperbolic equations (Burgers' and SWE) to challenge RandNet-Parareal on systems for which Parareal is known to struggle, %
with slow or non-convergent behavior~\cite{ariel2016parareal,bal2005convergence,dai2013stable,gander2007analysis,  staff2005stability}. Previous works have developed ad-hoc %
coarse solvers to address Parareal's slow convergence for Burgers'~\cite{chen2014use,jin2023learning, schmitt2018numerical, song2023analysis}, and for SWE~\cite{abel2020multigrid,haut2014asymptotic,nielsen2018communication,steinstraesser2024parallel}. Here, we adopt a different strategy: by leveraging the generalization capabilities of RandNets within %
the Parareal algorithm, we %
enhance the performance of standard, off-the-shelf integration methods such as Runge-Kutta, obtaining 
speed gains up to x125 and x22 compared to the accurate integrator $\f$ and Parareal, respectively. All experiments have been executed on Dell PowerEdge C6420 compute nodes each with 2 x Intel Xeon Platinum 826 (Cascade Lake) 2.9 GHz 24-core processors, 48 cores and 192 GB DDR4-2933 RAM per node. %
To illustrate our proposed algorithm and facilitate code adoption, we provide a step-by-step Jupyter notebook outlining RandNet-Parareal. Moreover, all simulation outcomes, including tables and figures, are fully reproducible and accompanied by the necessary Python code at \href{https://github.com/Parallel-in-Time-Differential-Equations/RandNet-Parareal}{https://github.com/Parallel-in-Time-Differential-Equations/RandNet-Parareal}.

It is well acknowledged that comparing PinT methods based on different working principles is extremely hard, with \cite{ong2020applications} representing a recent survey article with some comparisons. Quoting \cite{ong2020applications},``caution should be taken when directly comparing speedup numbers across methods and implementations. In particular, some of the speedup and efficiency numbers are only theoretical in nature, and many of the parallel time methods do not address the storage or communication overhead of the parallel time integrator''. \cite{Gander2023} is one of very few recent attempts to systematically compare different PinT classes. However, it is limited exclusively to the Dahlquist problem. Thus, it has become conventional to compare new techniques to the existing state-of-the-art methods within the same group of solvers. This is why, in this work, we compare RandNet-Parareal with the original Parareal and its recently improved versions, GParareal \cite{pentland2023gparareal}, and nnGParareal \cite{nngparareal}.

The rest of the paper is organized as follows. In Section~\ref{sec:para}, we describe the Parareal algorithm. Section~\ref{sec:nngpara} briefly explains GParareal and nnGParareal, focusing on the latter. RandNet-Parareal is introduced in Section~\ref{sec:elm}, while %
Sections~\ref{sec:experiments} and~\ref{sec:discussion} present our numerical results, and a final discussion. A computational complexity analysis of RandNet-Parareal, a robustness evaluation of the proposed algorithm, complementary simulation studies, and other additional results are available in the Supplementary Material.

\noindent {\bf Notation.} 
We denote by $\boldsymbol{v}\in\mathbb{R}^n$ a column vector with entries ${v}_i$, $i\in\{1,\ldots,n\}$, and by $\| \boldsymbol{v}\|  $ and $\| \boldsymbol{v}\|_{\infty}$ its Euclidean and infinity norms, respectively. We use $A\in \mathbb{R}^{n \times  m }$ to denote a real-valued $n\times m$ matrix, $n, m \in \mathbb{N}$, with elements $A _{ij} $, $j$th column $A_{(\cdot, j)}$, $j \in \{ 1, \dots m\}$, and $i$th row $A_{(i, \cdot)}$, $i \in \{ 1, \dots, n\}$. We write $A^\top$, $A^\dagger$,  and $\|A\|_{\rm F}$ for the $A$ matrix transpose, Moore-Penrose pseudoinverse, and  Frobenius norm, respectively. $\mathbb{I}_n$ denotes the identity matrix of dimension $n$.

\section{The Parareal algorithm}
\label{sec:para}

The idea of Parareal is to solve the $d$-dimensional ODE (and similarly PDE) system \eqref{eq:ode} in a parallel-in-time fashion, dividing the original IVP into $N$ sub-IVPs%
\[
\frac{d \boldsymbol{u}_i}{d t}=h\left(\boldsymbol{u}_i\left(t \mid \boldsymbol{U}_i\right), t\right), \quad  t \in\left[t_i, t_{i+1}\right], \quad \boldsymbol{u}_i\left(t_i\right)=\boldsymbol{U}_i, \quad \text{ for } \enspace i=0,\ldots,N-1,
\]
where the number of time intervals $N$ is also the number of available machines/cores/processors, 
$\boldsymbol{u}_i\left(t \mid \boldsymbol{U}_i\right)$ is the solution at time $t$ of the $i^{\rm th}$ IVP with initial condition $\boldsymbol{u}(t_i)=\boldsymbol{U}_i\in\mathbb{R}^d$, $  i=0,\ldots, N-1$. If the initial conditions were known and satisfied the continuity conditions $\boldsymbol{U}_i=\boldsymbol{u}_{i-1}\left(t_i|\boldsymbol{U}_{i-1}\right)$ (for the coherent temporal evolution of the system across sub-intervals), then the sub-IVPs could be %
trivially solved in parallel on a dedicated machine. Unfortunately, this is not the case, as only the first initial condition $\boldsymbol{U}_0 =\boldsymbol{u}^0\in \mathbb{R}^d$ at time $t_0$ appears available. 
To account for this, Parareal introduces another numerical integrator $\g$, much faster but less accurate than $\f$, %
to approximate the missing initial conditions $\bs{U}_i$, $i=1, \ldots, N-1$, \textit{sequentially}. $\g$ trades off accuracy for computational feasibility, usually taking seconds/minutes instead of hours/days of $\f$\footnote{$\f$ and $\g$ can be two different solvers or the same solver with different time steps.}.

The algorithm works as follows. 
We use $\bs{U}_i^k$ to denote the Parareal approximation of $\boldsymbol{u}_i(t_i)=\boldsymbol{U}_i$ %
at iteration $k\geq 0$. At $k=0$, the initial conditions $\{\boldsymbol{U}_{i}^{0}\}_{i=1}^{N-1}$ are initialized using a \textit{sequential} application of the coarse solver $\g$, obtaining $\boldsymbol{U}_i^0=\g(\boldsymbol{U}_{i-1}^0)$, $i=1,\ldots, N-1$, with $\boldsymbol{U}_0^0=\boldsymbol{U}_0$. At $k \geq 1$,
the obtained initial conditions $\boldsymbol{U}_{i-1}^{k-1}$ are \lq\lq propagated\rq\rq\ through $\f$ in \textit{parallel} on $N$ cores to obtain %
$\f(\boldsymbol{U}_{i-1}^{k-1})$, $i=1,\ldots, N$.
Note that for every initial condition $\boldsymbol{U}_{i-1}^{k-1}$,
we compute both $\f(\boldsymbol{U}_{i-1}^{k-1})$, i.e. a precise evaluation of $\bs{u}_{i-1}(t_i|U_{i-1}^{k-1})$, and $\g(\boldsymbol{U}_{i-1}^{k-1})$, an inaccurate evaluation of the same term. Hence,  we can interpret $\f$ and $\g$ as functions mapping
an initial condition to the next one, thereby evolving \eqref{eq:ode} by %
one interval. %
We can then use their difference, $(\f-\g)(\boldsymbol{U}_{i-1}^{k-1})$, to correct the inaccuracy of $\g$ on future evaluations. This gives rise to the original Parareal predictor-corrector rule $\boldsymbol{U}_i^{k}=\mathscr{G}(\boldsymbol{U}_{i-1}^{k})+(\mathscr{F}-\mathscr{G})(\boldsymbol{U}_{i-1}^{k-1})$,
with $i=1, \ldots, N-1$, $k\geq 1$ ~\cite{gander2007analysis},  where the \textit{sequential} prediction $\mathscr{G}(\boldsymbol{U}_{i-1}^{k})$ is corrected by adding the discrepancy $\f-\g$ computed at the previous iteration $k-1$. However, this formulation can be changed to use data from the current iteration $k$~\cite{pentland2023gparareal}, and generalized to account for different ways of computing the discrepancy, leading to~\cite{nngparareal} 
\begin{equation}
    \boldsymbol{U}_i^k = \g(\boldsymbol{U}_{i-1}^{k}) + \fhat(\boldsymbol{U}_{i-1}^{k}),
    \label{eq:update_rule_generic}
\end{equation}
where $\fhat: \R^d \rightarrow \R^d$ specifies how the correction function $\f-\g$ is computed or approximated based on some observation $\boldsymbol{U} \in \R^d$. 
Parareal uses 
\begin{equation}\label{fpara}
\fhat_{\rm Para}(\boldsymbol{U}_{i-1}^{k})=(\f-\g)(\boldsymbol{U}_{i-1}^{k-1}),
\end{equation}
while other variants will be introduced in the subsequent sections. The Parareal solution \eqref{eq:update_rule_generic} is considered converged for a given threshold $\epsilon>0$ and up to time $t_L \leq t_N$, 
if  solutions across consecutive iterations have stabilized. That is, for some pre-defined accuracy level $\epsilon>0$, it holds that
\begin{equation}
    \|\boldsymbol{U}^k_i-\boldsymbol{U}^{k-1}_i\|_{\infty} < \epsilon,\quad  0<i\leq L\leq N-1.
    \label{eq:stop_crit}
\end{equation}
Other stopping criteria are also possible~\citep{samaddar2010parallelization, samaddar2019application}. Converged Parareal approximations $\boldsymbol{U}_i^k$, $i\leq L$, are no longer iterated to avoid unnecessary overhead~\citep{elwasif2011dependency,garrido2006convergent,nngparareal, pentland2023gparareal,pentland2022stochastic}. Instead, unconverged solution values $\boldsymbol{U}_i^k$, $i>L$, are updated during future iterations by first running $\f$ in parallel and then using the prediction-correction rule \eqref{eq:update_rule_generic}. The Parareal algorithm stops at some iteration $K_{\rm Para} \leq N$ when all initial conditions have converged, that is when \eqref{eq:stop_crit} is satisfied with $L=N-1$ %
and thus $K_{\rm Para}=k$.  
Note that during every Parareal iteration $k>1$, the ``leftmost'' fine solver evaluation $\f(\bs{U}_L^k)$ is either run from the outcome of a previous fine computation $\bs{U}_L^k=\f(\bs{U}_{L-1}^{k-1})$, or from a converged initial condition $\|\boldsymbol{U}^k_L-\boldsymbol{U}^{k-1}_L\|_{\infty} < \epsilon$. This guarantees that, either way, the maximum number of iterations to convergence for \textit{any} Parareal-based algorithm is $K_{\rm Para}=N$, in which case it sequentially attains the fine solver solution, with the added computational cost of running $\g$ and evaluating $\widehat f$ $N$ times. A Parareal pseudocode is presented in Algorithm~\ref{alg:para} in Supplementary Material~\ref{supp:psudocodes}. 

\section{GParareal and Nearest Neighbors GParareal}
\label{sec:nngpara}
The performance of Parareal can be improved by a careful selection of $\fhat$ in \eqref{eq:update_rule_generic}, 
combined with a better use of the available information present at iteration $k$. Let $\mathcal{D}_k$ denote the dataset consisting of $Nk$ pairs of inputs $\boldsymbol{U}_{i-1}^j \in \mathbb{R}^d$ and their corresponding outputs $(\f-\g)(\boldsymbol{U}_{i-1}^j)\in \mathbb{R}^d$, $i=1,\ldots, N$, $j=0,\ldots, k-1$, that is
\begin{equation}\label{Dk}
\mathcal{D}_k:= \{(\boldsymbol{U}_{i-1}^{j}, (\f-\g)(\boldsymbol{U}_{i-1}^{j})), \enspace i=1,\ldots,N, \enspace j=0,\ldots,k-1\}.
\end{equation}
While Parareal relies on one observation to construct the correction $\fhat$ in \eqref{fpara}, GParareal and following works, including this one, use all the discrepancy terms $\f-\g$ and information in $\mathcal{D}_k$ to make their predictions.  
The idea of GParareal is to learn the map $\mathbb{R}^d\rightarrow \mathbb{R}^d$, $\boldsymbol U_{i-1}^k\mapsto (\f-\g)(\boldsymbol U_{i-1}^k)$, via $d$ independent scalar GPs $\mathbb{R}^d\rightarrow \mathbb{R}$, $\boldsymbol U_{i-1}^k\mapsto \widehat f_{{\rm GPara}}^{(s)}(\boldsymbol U_{i-1}^k)$, $s=1,\ldots, d$, one per ODE dimension, whose predictions are  concatenated into $\fhat_{\rm GPara}(\boldsymbol{U}_{i-1}^k)=(\fhat_{{\rm GPara}}^{(1)}(\boldsymbol{U}_{i-1}^k),\ldots, \fhat_{{\rm GPara}}^{(d)}(\boldsymbol{U}_{i-1}^k))^\top\in\mathbb{R}^d$, and finally plugged into the predictor-corrector rule \eqref{eq:update_rule_generic}. In particular, 
 each GP prediction $\fhat_{\rm GPara}^{(s)}(\boldsymbol{U}_{i-1}^k)$ is obtained as the GP posterior mean $\mu_{\mathcal{D}_k}^{(s)}(\boldsymbol{U}_{i-1}^k) \in \mathbb{R}$, computed by conditioning the corresponding GP prior on the dataset $\mathcal{D}_k$, i.e. $\fhat_{\rm GPara}^{(s)}(\boldsymbol{U}_{i-1}^k) = \mu^{(s)}_{\mathcal{D}_{k}}(\boldsymbol{U}_{i-1}^k)$. We refer to Supplementary Material \ref{supp:nngp_full_form} and~\cite{pentland2023gparareal} for a thorough description of the algorithm, including all relevant quantities of interest, namely the $d$ GP priors, the likelihood, the hyperparameters and their optimization procedure, and an explicit expression of the posterior means. %
Here, it is worth highlighting that the GPs are trained once per iteration to leverage the new incoming data, and then their predictions are used to \textit{sequentially} update the initial conditions in \eqref{eq:update_rule_generic}. %
Using all information stored in $\mathcal{D}_k$ instead of a single observation (as for Parareal) is the primary driver of faster convergence rates experienced by GParareal. Other benefits of this algorithm are increased stability to different initial conditions, the ability to incorporate legacy data (that is, the possibility of using datasets  coming from previous runs of the algorithm with different starting conditions or settings, leading to faster convergence), lower sensitivity to poor choices of the coarse solver $\g$, and the possibility of parallelizing the training of the $d$ GPs %
over the $N$ available cores. The main drawback of GParareal is  the heavy computational burden incurred when inverting the GP covariance matrices, which is of order $O(d(Nk)^3)$ at iteration $k$. This negatively impacts the algorithm's wallclock time, which may be higher than Parareal despite a lower number of iterations needed to converge. This is why GParareal has been proposed  mainly for low-dimensional ODE systems with a relatively small number of processors/intervals $N$ (up to hundreds), limiting its use and parallel scalability~\cite{pentland2023gparareal}.%

The nnGParareal algorithm \cite{nngparareal} has been proposed to tackle GParareal's scalability issue, sensibly reducing the computational time and memory footprint of GPs by using their nns  version (nnGPs). In this framework, at iteration $k$, the $d$ GPs are all trained on a smaller dataset of size $m$, $\mathcal{D}_{i-1,k}$, composed out of the $m$ nns (in Euclidean distance) of $\bs{U}_{i-1}^k$ in $\mathcal{D}_k$, leading to the nnGParareal correction $\fhat_{\rm nnGPara}(\boldsymbol U_{i-1}^k)=(\fhat_{\rm nnGPara}^{(1)}(\boldsymbol U_{i-1}^k),\ldots,\fhat_{\rm nnGPara}^{(d)}(\boldsymbol U_{i-1}^k))^\top$, with
\[
\fhat_{\rm nnGPara}^{(s)}(\boldsymbol{U}_{i-1}^k) = \mu^{(s)}_{\mathcal{D}_{i-1,k}}(\boldsymbol{U}_{i-1}^k), \quad s=1,\ldots, d.
\]
Here, $\mu^{(s)}_{\mathcal{D}_{i-1,k}}\in\mathbb{R}$, $s=1,\ldots, d$, denotes the nnGP posterior mean computed by conditioning the corresponding GP prior on the reduced dataset $D_{i-1,k}$ of size $m$.
Due to the decreased sample size, each nnGP covariance matrix can be inverted at a cost of $O(m^3)$ independent of $k$ or $N$. However, contrary to GParareal which trains the GPs once per iteration, the nnGPs are re-trained \textit{every time a new prediction $\widehat f_{\rm nnGPara}(\boldsymbol U_{i-1}^k)$} is made, which are at most $N-k$ at iteration $k$ (as at least $k$ intervals have converged at iteration $k$), yielding a combined $O(d(N-k)m^3)$ complexity. Several experiments on different ODE and PDE systems have shown that $m\in\{15,\ldots,20\}$ offer accuracy
comparable to the full GP~\cite{nngparareal} at a much lower cost. Although faster than GParareal, nnGParareal still exhibits some of the drawbacks inherited from the GP framework, such as the cost of optimizing the hyperparameters through a numerical maximization of a non-convex likelihood, and the use of $d$ scalar nnGPs. The latter is particularly critical. On the one hand,  despite the possibility of training the $d$ nnGPs in parallel, %
the inversion of a $m \times m$ matrix is so efficient that parallel overheads may outweigh the theoretical benefits. On the other hand, when solving PDEs, nnGParareal %
will incur additional costs due to insufficient hardware resources, as usually $d \gg N$, forcing the $d$ nnGPs to queue among the $N$ available processors, which is why the algorithm has been proposed for high-dimensional ODE and PDE systems with $d\leq N$. We refer to Supplementary Material~\ref{supp:nngp_full_form} and~\cite{nngparareal} for more details on nnGParareal, and to Algorithm~\ref{alg:nngp} in  Supplementary Material~\ref{supp:psudocodes} for the pseudocode of the nnGP training.  In the next section, we address the nnGParareal issues by introducing RandNets.

\section{Random neural networks Parareal (RandNets-Parareal)}
\label{sec:elm}

In RandNet-Parareal, we propose to learn the map  $\mathbb{R}^d\to\mathbb{R}^d$, $\bs U\mapsto(\f-\g)(\bs U)$ via RandNets, obtaining the RandNet-Parareal correction $\widehat f_{\rm RandNet\text{-}Para}$, which we then use within the predictor-corrector rule \eqref{eq:update_rule_generic}. Prior to that, we define how RandNets work in a general setting with input $\bs U\in\mathbb{R}^d$ and output or target $\bs Y\in\mathbb{R}^d$. Later in the text we will go back to the input of interest $\bs{U}_i^k$.
Let $M$ denote the number of hidden neurons, %
and $H_{W}^{A,\bs{\zeta}}(\bs{U})$ be a single-hidden-layer feed-forward neural network used to learn $\f-\g$, given by
\begin{equation}\label{HA}
H_{W}^{A,\bs{\zeta}}(\bs{U})=W^\top\bs{\sigma}(A\bs{U}+\bs{\zeta}) \in \mathbb{R}^d, \quad \bs{U} \in \mathbb{R}^d,
\end{equation}
where $A\in\mathbb{R}^{M\times d}$ is the matrix of random, non-trainable weights of the hidden layer, $\bs{\zeta}\in \mathbb{R}^{M}$ is a random non-trainable bias vector, and $W\in \mathbb{R}^{M\times d}$ is the matrix of trainable output weights.  Here, $\bs{\sigma}: \mathbb{R}^M \rightarrow \mathbb{R}^M$ denotes an activation function obtained as the componentwise application of a non-linear map $\sigma: \mathbb{R} \rightarrow \mathbb{R}$ which we choose to be ReLU $\sigma(x) = \max(x,0)$ with $x\in \mathbb{R}$, to satisfy the assumption of Proposition \ref{prop:approc_bound} below. %
The entries of $A$ and  $\bs{\zeta}$ are randomly sampled from given distributions $\mathcal{P}_A$ and $\mathcal{P}_{\bs{\zeta}}$, respectively, and kept fixed. After observing the dataset $\mathcal{D}_k$, the output weights $W$ are obtained as the minimum $\ell_2$ norm least squares (or simply min-norm least squares) estimator or as the solution of the following penalized empirical minimization problem: %
\[
\widehat{W}^{\mathcal{D}_k} = \lim_{\lambda\rightarrow 0} \arg \min_{W \in \mathbb{R}^{M\times d}} \left\{ \sum_{(\bs{U}, \bs{Y})\in\mathcal{D}_k} \left \|H_W^{A,\bs{\zeta}}(\bs{U})-\bs{Y} \right \|^2 + \lambda \left \| W \right\|_{\rm F}^2 \right\},
\]
which is also called a ``ridgeless'' (interpolation) estimator~\cite{ridgeless2022}, and can be more compactly written as  %
\begin{equation} 
\label{ridgeless}
\widehat{W}^{\mathcal{D}_k}=\lim_{\lambda\rightarrow 0} \left( X^\top X +\lambda \mathbb{I}_{M}\right)^{-1} X^\top Y.
\end{equation}
Here, $X\in\mathbb{R}^{Nk\times M}$ is a matrix with $(X_{(l,\cdot)})^\top:=\bs{\sigma}(A(U_{(l,\cdot)})^\top+\bs{\zeta})$, $l=1,\ldots,Nk$, and $U, Y\in \mathbb{R}^{Nk\times d}$ are the collection of inputs and outputs of $\mathcal{D}_k$ in matrix form,respectively, defined as $(U_{(l,\cdot)})^\top=\bs U_i^j$, $(Y_{(l,\cdot)})^\top=\bs Y_i^j$, $l=jN+i+1$, $i=0,\ldots, N-1$, $j=0,\ldots, k-1$. Whenever $Nk\geq M$ and the rank of $X^\top X\in \mathbb{R}^{M\times M}$ is $M$, \eqref{ridgeless} reduces to the standard least squares estimator $\widehat{W}^{\mathcal{D}_k}=\left( X^\top X \right)^{-1} X^\top Y$, while if the rank of $X^\top X$ is $Nk$, the solution admits a closed form
\begin{equation*}\label{What}\widehat{W}^{\mathcal{D}_k}=\left( X^\top X \right)^\dagger X^\top Y.
\end{equation*}

We get inspired by \cite{nngparareal}, where only $m$ nns are used in the training. In this setting, $M\gg Nk=m $, and in this overparametrized linear regression case, the ridgeless estimator interpolates the training data, which is a desirable feature since the problem is genuinely deterministic \cite{han2023distribution,meimontanari2022}.

Several ingredients control the performance of RandNets, such as the dimension of the network $M$ and the choice of distributions  $\mathcal{P}_A$ and $ \mathcal{P}_{\bs{\zeta}}$. In this work, we take the rows of the weight matrix $A$ and the bias entries of $\bs{\zeta}$ to be independent and uniformly distributed. For this case, the approximation bounds are available 
\cite[Proposition 3]{gonon2023approximation}, which we report below using our notation. %

\begin{proposition}[Approximation bound,~\cite{gonon2023approximation}, Proposition 3]
\label{prop:approc_bound}
    Let $H^*: \mathbb{R}^d \rightarrow \mathbb{R}$, $\bs{U} \longmapsto H^*(\bs{U})$ be an unknown function we wish to approximate with $H_W^{A,\bs{\zeta}}$ defined in \eqref{HA}. Suppose $H^*$ can be represented as $H^*(\bs{U})=\int_{\mathbb{R}^d} e^{i\langle\mathbf{w}, \bs{U}\rangle} g(\mathbf{w}) \mathrm{d} \mathbf{w}$
for some complex-valued function $g$ on $\mathbb{R}^d$ and all $\bs{U} \in \mathbb{R}^d$ with $\|\bs{U}\| \leq Q$, where $\langle \cdot, \cdot \rangle$ is the inner product on $\R^d$. Assume that $\int_{\mathbb{R}^d} \max \left(1,\|\mathbf{w}\|^{2 d+6}\right)|g(\mathbf{w})|^2 \mathrm{~d} \mathbf{w}<\infty$. %
For $\rho>0$, suppose the rows of $A$ are i.i.d. random variables with uniform distribution on $B_\rho \subset \mathbb{R}^d$, the Euclidean ball of radius $\rho$ around $\mathbf{0}$, %
and that the $M$ components of $\bs{\zeta}$ are i.i.d. uniform random variables on $[-\max (Q \rho, 1), \max (Q \rho, 1)]$. Assume that $A$ and $\bs{\zeta}$ are independent and let $\sigma: \mathbb{R} \rightarrow$ $\mathbb{R}$ be given by $\sigma(x)=\max (x, 0)$. Then, there exist a $\mathbb{R}^{M\times d}$-valued random variable $W$ and an explicit (see (33) in \cite{gonon2023approximation}) constant $C^*>0$ such that
\[
\mathbb{E}\left[\|H_{W}^{A, \boldsymbol{\zeta}}(\bs{U})-H^*(\bs{U})\|^2\right] \leq \tfrac{C^*}{M},
\]
and for any $\delta \in(0,1)$, the random neural network $H_{{W}}^{{A}, \boldsymbol{\zeta}}$ satisfies
$$ \mathbb{P}\Big(
\Big(\int_{\mathbb{R}^d}\|H_{{W}}^{{A}, \boldsymbol{\zeta}}(\bs{U})-H^*(\bs{U})\|^2 \mu_{\bs{U}}(\mathrm{d} \bs{U})\Big)^{1 / 2} \leq \tfrac{\sqrt{C^*}}{\delta \sqrt{M}}\Big)\geq 1-\delta.
$$
\end{proposition}

Our choice of $\mathcal{P}_A$ and $ \mathcal{P}_{\bs{\zeta}}$ satisfies the conditions of Proposition~\ref{prop:approc_bound} if  $\|\bs{U}\| \leq Q$. If this is not met, 
we rescale the ODE/PDE system via a change of variables. We found these bounds empirically useful in informing a good choice for the sampling distribution, which we follow. If no prior information were available, the common approach would have been to take $\mathcal{P}_A\sim \text{Unif}(-a,a)^{M \times d} $, $\mathcal{P}_{\bs{\zeta}}\sim \text{Unif}(-b,b)^M$, and optimize $a,b\in \mathbb{R}^+$ via  expensive cross-validation procedure. 

Unlike nnGParareal, GParareal, and the corresponding nnGPs and GPs, training RandNets is so fast that parallelization across the $d$ dimensions is unnecessary. Hence, the predictions of the random network are computed jointly on all $d$ coordinates, yielding the RandNet-Parareal correction function
\begin{equation}\label{eq:randnet}    
\fhat_{\rm RandNet\text{-}Para}(\bs{U}_{i-1}^k)=H_{\widehat{W}^{\mathcal{D}_{i-1,k}}}^{A, \bs{\zeta}}(\bs{U}_{i-1}^k)\in \mathbb{R}^d.
\end{equation}
Here, the estimated weights $\widehat{W}^{\mathcal{D}_{i-1,k}}$ are obtained using the reduced dataset $\mathcal{D}_{i-1,k}$ consisting of the $m_{\rm RandNet}$ nns of $\bs{U}_{i-1}^k$, requiring the retraining of the RandNet for every prediction. Employing a multi-output model instead of independently training $d$ scalar-output
models addresses one of the %
pitfalls of GPs%
, allowing for better scalability when $d \gg N$. The fact that training the RandNets reduces to a closed-form ridgeless interpolation solution  presents a substantial difference and improvement with respect to  %
(nn)GPs. Moreover, expensive hyperparameter optimization is avoided in RandNets, addressing the other major pitfall of GParareal and nnGParareal. The pseudocode for training RandNets is reported in Algorithm~\ref{alg:RandNets}  in  Supplementary Material~\ref{supp:psudocodes}. 

In Supplementary Material~\ref{supp:comp_complx}, we derive the theoretical computational costs of nnGParareal and RandNet-Parareal, illustrating them as a function of dimension $d$ and number of processors $N$ in Figure \ref{fig:comp_complx}. These theoretical findings confirm the significantly superior scalability of RandNet-Parareal which we observe in the numerical experiments reported in Section \ref{sec:experiments}.
In Supplementary Material~\ref{supp:robustness}, we study the robustness of RandNet-Parareal to changes in the number of nns $m_{\rm RandNet}$ (and thus the input data size), the number of neurons $M$, and the randomly sampled network weights $A, \bs{\zeta}$. %
Intuitively, one might anticipate that a larger data sample would yield a more accurate approximation of the correction  $\f-\g$, and that a higher number of neurons $M$ would reduce the prediction error of RandNets (as in Proposition \ref{prop:approc_bound}). %
One may also suspect the algorithm to be sensitive %
to the particular sampling seed. %
Remarkably, our empirical findings demonstrate that these factors have a limited impact on the number of iterations needed by RandNet-Parareal to converge, which remains largely consistent (up to a few iterations) across different values and ODE/PDE systems, for sensible choices of $m_{\rm RandNet}$ and $M$. For the end user, this eliminates the need of ad-hoc tuning, making the proposed RandNet-Parareal a convenient out-of-the-box algorithm.

\section{Numerical Experiments}
\label{sec:experiments}

In this section, we first compare the performance of Parareal, nnGParareal, and RandNet-Parareal on the viscous Burgers' equation (one spatial dimension and one variable, also considered in nnGParareal~\cite{nngparareal}),  to showcase Parareal and nnGParareal challenges %
as the number of space discretization and, correspondingly, the dimensions $d$, increases. Then, we consider the Diffusion-Reaction equation, a larger system %
defined on a two-dimensional spatial domain with two 
non-linearly coupled variables, and the SWEs (two spatial dimensions and three variables), representing a suitable framework for modeling free-surface flow problems on a two-dimensional domain. Two additional challenging systems, the 2D and 3D Brusselator PDEs, known for their complex behavior, including oscillations, spatial patterns, and chaos, are considered in Supplementary Material \ref{supp:brusselator}. 
The simulation setups used for obtaining the results in this section are provided in Supplementary Material \ref{supp:sim_setup}, with the corresponding accuracies and runtimes for RandNet-Parareal, Parareal, and nnGParareal reported in Supplementary Material \ref{supp:accuracy}.

Let $T_\f$ and $T_\g$ be the time it takes to run $\f$ and $\g$ over one interval $[t_i, t_{i+1}]$, respectively, and let $N_\f$ and $N_\g$ denote the number of steps for the fine and coarse solvers over one interval, respectively. We can measure the parallel efficiency of an algorithm via its parallel speed-up $S_{\rm alg}$, defined as the ratio of the serial over the parallel runtime, i.e. $S_{\rm alg}:=N T_\f/T_{\rm alg}$. $S_{\rm alg}$ captures the wallclock gains of parallel procedures and, unlike  other quantities (such as the number of algorithm iterations needed  to converge),  also includes the model training cost.  

\subsection{Viscous Burgers' equation}
\label{sec:burgers}
Our initial example is a non-linear, one-dimensional PDE (illustrated in Figure~\ref{fig:burgers_system_evolution} of Supplementary Material \ref{supp:illustrations}) %
exhibiting hyperbolic behavior~\cite{schmitt2018numerical}, described by the equation
\begin{equation}
    v_t = \nu v_{xx} - vv_x, \quad (x,t)\in [-L,L]%
    \times [t_0,t_N],
    \label{eq:burg}
\end{equation}
with initial condition $v(x, t_0)=v_0(x)$, $x\in [-L,L], L>0$, and Dirichlet boundary conditions $v(-L,t)=v(L,t)$, $v_x(-L,t)=v_x(L,t)$, $t \in [t_0, t_N]$. We use the same setting and parameter values as in~\cite{nngparareal}. More specifically, we choose $L=1$, diffusion coefficient $\nu=0.01$, and discretize the spatial domain using finite difference~\cite{fornberg1988generation} and equally spaced points $x_{j+1}=x_j+\Delta x$, with $\Delta x = 2L/d$ and $j=0,\ldots,d$. We hence reformulate the PDE as a $d$-dimensional ODE system. 

In our first numerical experiment, we choose $N=d=128$, $v_0(x)=0.5(\cos (\frac{9}{2}\pi x)+1)$, $t_0=0$, and $t_N=5.9$ as in~\cite{nngparareal}, and consider $\g={\rm RK1}, \f= {\rm RK8}$, $N_\g=4$ and $N_\f=4e^4$, where ${\rm RK1}$ stands for Runge-Kutta  of order 1, and similarly for ${\rm RK4}$ and ${\rm RK8}$.  The results, reported at the top of Table \ref{tab:burgers_summ}, show how RandNet-Parareal converges in fewer iterations and has a higher speed-up than Parareal and nnGParareal. The difference in the model training costs is striking, with the nnGP's being approximately 700 times higher than that of RandNets, reducing thus its potential speed-up.

As real-world (one-dimensional) problems would require a higher spatial discretization, we increase $d$ by one thousand to $d=1128$, keeping $N$ fixed. Unlike assuming matching hardware resources to the system size (as implicitly done in~\cite{nngparareal}, where $d=N$), we deliberately do not increase $N$ to assess the algorithms' performances under constrained conditions. %
Instead, 
both time discretization numbers are increased to $N_\f=6e^5$ and $N_\g=293$ (resulting thus in longer $T_\f$ and $T_\g$ times) to account for the finer spatial mesh~\cite{leveque2007finite}. As observed from the bottom of Table \ref{tab:burgers_summ}, as $d/N>1$, nnGParareal's issues become more pronounced, as the $d$ scalar GPs cannot be  run all in parallel across the $N$ processors, but need $d/N=10$ runs instead, slowing down the algorithm. In contrast, RandNet-Parareal has a training cost comparable with the previous example, leading to an even higher speed-up, running in approximately 38 minutes compared to the almost 13 hours of Parareal.%

\begin{table}
    \footnotesize
    \caption{Empirical scalability and speed-up analysis for viscous Burgers' equation}
    \caption*{$d=128$, $N=128$}
        \label{tab:burgers_summ}
    \centering
\begin{tabular}{lcccC{1.1cm}C{1.3cm}c}
\toprule
Algorithm & $K$ & $NT_{\g}$ & $T_{\f}$ & $T_{\rm model}$ & $T_{\rm alg}$ & $S_{\rm alg}$\\
\midrule
Fine & -- & -- & -- & -- & 13h 5m & 1\\
Parareal & 90 & 0s & 6m & 0s & 8h 54m & 1.47\\
nnGParareal & 14 & 0s & 6m & 12m & 1h 39m & 7.90\\
RandNet-Parareal & 10 & 0s & 6m & 1s & 1h 2m & \textbf{12.61}\\
\bottomrule
\vspace{-0.5cm}
\end{tabular}
\caption*{$d=1128$, $N=128$}
\begin{tabular}{lcccC{1.1cm}C{1.3cm}c}
\toprule
Algorithm & $K$ & $NT_{\g}$ & $T_{\f}$ & $T_{\rm model}$ & $T_{\rm alg}$ & $S_{\rm alg}$\\
\midrule
Fine & -- & -- & -- & -- & 18h 52m & 1\\
Parareal & 91 & 0s & 9m & 0s & 12h 57m & 1.41\\
nnGParareal & 6 & 2s & 9m & 1h 25m & 2h 17m & 8.26\\
RandNet-Parareal & 4 & 2s & 9m & 1s & 38m & \textbf{29.98}\\
\bottomrule
\end{tabular}\\ 
    \caption*{Speed-up $S_{\rm alg}$ of Parareal, nnGParareal ($m_{\rm nnGP}\text{=}18$), and RandNet-Parareal ($m_{\rm RandNet}\text{=}4$, $M\text{=}100$) for the 1D viscous Burgers' equation. $T_\f$ and $T_\g$ are the interval runtimes of the fine and coarse solvers, respectively, $K$ the number of iterations to converge, $T_{\rm model}$ the overall time to evaluate $\fhat$ across $K$ iterations, including training and predicting, and $T_{\rm alg}$ thealgorithm runtime.%
    }
\end{table}

\subsection{Diffusion-Reaction system}
\label{sec:diffreact}

We now turn to a more challenging case study. The Diffusion-Reaction equation~\cite{takamoto2022pdebench} (illustrated in Figure~\ref{fig:diffreact_system_evolution} in Supplementary Material~\ref{supp:illustrations}) is a system of two non-linearly coupled variables, the activator $u=u(t, x, y)$ and the inhibitor $v=v(t, x, y)$, defined on a two-dimensional spatial domain as
\[
\partial_t u=D_u \partial_{x x} u+D_u \partial_{y y} u+R_u, \quad \partial_t v=D_v \partial_{x x} v+D_v \partial_{y y} v+R_v.
\]
Here, $D_u$, $D_v$ are the diffusion coefficients for the activator and inhibitor, respectively, and $R_u=$ $R_u(u, v)$,  $R_v=R_v(u, v)$ are their reaction functions defined by the Fitzhugh-Nagumo equation~\cite{klaasen1984stationary}
\[
\begin{aligned}
& R_u(u, v)=u-u^3-c-v, \quad
& R_v(u, v)=u-v,
\end{aligned}
\]
where $c=5e^{-3}$, $D_u=1e^{-3}$, and $D_v=5e^{-3}$. We take $(x,y) \in (-1,1)^2$ and $t \in [0, 20]$. The initial condition $u(0,x,y)$ is generated as standard Gaussian noise.  We apply a no-flow Neumann boundary condition $D_u \partial_x u=0$, $D_v \partial_x v=0$, $D_u \partial_y u=0$, $D_v \partial_y v=0$ for $(x, y) \in(-1,1)^2$. The spatial domain is discretized by the finite volume method~\cite{darwish2016finite}, resulting in a $d=2N_x N_y$-dimensional ODE with $N_x$ and $N_y$ the number of space discretizations along $x$ and $y$, respectively. The time integration is conducted with RK of variable order for $\g$ and $\f$  (see Table \ref{tab:diffreact_summ_setup} in  Supplementary Material~\ref{supp:sim_setup}).

\begin{figure}[h]
     \centering     \includegraphics[width=0.95\linewidth]{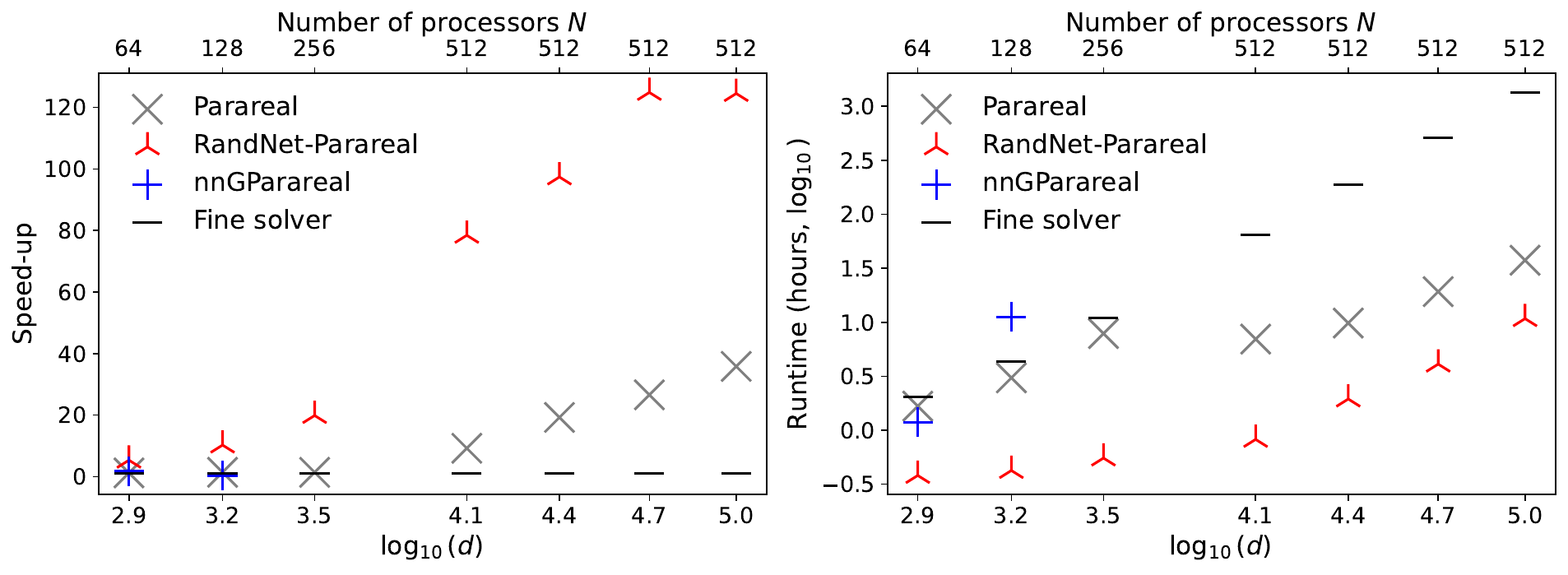} %
     \caption{Speed-ups (left) and runtimes (right) of Parareal, nnGParareal ($m_{\rm nnGP}$=$20$), and RandNet-Parareal ($m_{\rm RandNet}\text{=}4$, $M\text{=}100$) for the two-dimensional Diffusion-Reaction system versus the number $d$ of dimensions  (bottom x-axis) and $N$ cores (top x-axis) capped at $512$ to simulate limited resources.}
     \label{fig:diff_react_speedup}
 \end{figure}

As in the previous example, we conduct two experiments for this system, with speed-ups and runtimes reported in Figure~\ref{fig:diff_react_speedup}. In the first one, we increased $d$ and $N$ proportionately (with $d/N\in [11,13]$) while maintaining all other quantities (i.e. $\g,\f, m_{\rm nnGP}, m_{\rm RandNet}$) fixed until $N=256$. This scenario reflects a situation where more resources are allocated to solve larger problem sizes. In contrast, in the second experiment, $N$ remains fixed at $512$, with $d$ increasing proportionately with $N_\g$ to maintain algorithm stability. Moreover, $\f$ is chosen to be RK8, with $N_\f$ automatically selected by the used Python library \textit{scipy}~\cite{virtanen2020scipy}. %
This second setting simulates a scenario with constrained resources, where the user aims to solve the system using a finer spatial mesh. Table~\ref{tab:diffreact_speedup} in  Supplementary Material~\ref{supp:DiffReactTable} shows that for $N \geq 256$ and $d/N\gg 1$, nnGParareal fails to converge within a 48-hour budget. Parareal converges always, albeit at a considerably slower rate than RandNet-Parareal, which is x3-5 faster than Parareal (and up to x120 than the fine solver).
\subsection{Shallow water equation}
\label{sec:swe}
Finally, we focus on SWEs %
on a two-dimensional domain,  described by a system of hyperbolic PDEs
\[
\partial_t h+\nabla h \mathbf{u}=0, \quad \partial_t h \mathbf{u}+\nabla(u^2 h+\tfrac{1}{2} g_r h^2)=-g_r h \nabla b,
\]
where $\mathbf{u}=(u, v)$ represents the velocities in the horizontal $u=u(t,x,y)$ and vertical $v=v(t,x,y)$ directions, $h=h(t,x,y)$ denotes the water depth, $b=b(x,y)$ describes a (given) spatially varying bathymetry, and $h \mathbf{u}$ can be interpreted as the directional momentum components. The parameter $g_r$ describes the gravitational acceleration, while $\partial_t f$ denotes the partial derivative with respect to time, and $\nabla f$ the gradient of a function $f$.   Following~\cite{takamoto2022pdebench}, we solve a radial dam break scenario where a Gaussian-shaped water column (blue) inundates nearby plains (green) within a rectangular box subject to Neumann boundary conditions, causing the water to rebound off the sides of the box, as depicted in Figure~\ref{fig:swe_system_evolution}. More details on the simulation setup are given in Supplementary Material \ref{app:SWE}. 

In this case, our algorithm also converges much faster  than Parareal, with a
speed gain of x1.3-3.6, while nnGParareal fails to converge within the 48-hour time budget as $d\gg N$. %
Although the speed gain is lower than for the Diffusion-Reaction, the improvements are  remarkable. RandNet-Parareal takes up to 4-10 hours and 37 days less than the Parareal and sequential solver, respectively. 

\begin{figure}[ht!]
     \centering     \includegraphics[width=0.8\linewidth]{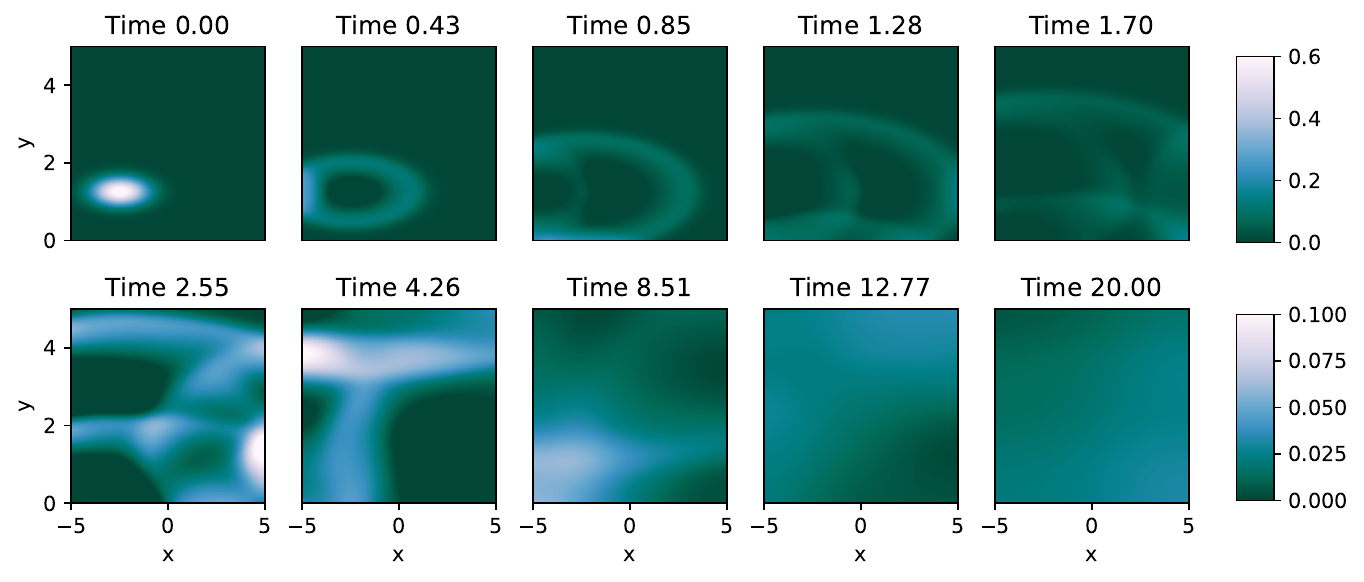}
     \caption{Numerical solution of the SWE for $(x,y) \in [-5,5]\times[0,5]$ with $N_x=264$ and $N_y=133$ for a range of system times $t$. Only the water depth $h$ (blue) is plotted.}
     \label{fig:swe_system_evolution}
 \end{figure}

\begin{table}
    \centering \footnotesize
\caption{Speed-up analysis for the shallow water PDE  as a $d$-dimensional ODE system, $N=235$ }
\begin{tabular}{cccccccc}
\toprule
$d$ & $K_{\rm Para}$ &$K_{\rm \scriptsize RandNet\text{-}Para}$ & $T_{\f}$ &$T_{\rm Para}$ &$T_{\rm RandNet\text{-}Para}$&$S_{\rm Para}$&$S_{\rm RandNet\text{-}Para}$ \\
\midrule
{15453}&52&14&22h 54m&5h 8m&1h 25m&4.47&\textbf{16.16}\\
{31104}&50&13&3d 2h&15h 43m&4h 9m&4.68&\textbf{17.69}\\
{60903}&14&9&13d 15h&19h 30m&12h 34m&16.73&\textbf{25.92}\\
{105336}&8&6&38d 4h&1d 7h&23h 34m&29.37&\textbf{38.90}\\
\bottomrule
\end{tabular}\\   
    \caption*{$K_\cdot$ is the number of iterations to converge, $T_\cdot$ wallclock time  and $S_\cdot$ speed-up  for the Parareal (Para) and RandNet-Parareal ($m_{\rm RandNet}\text{=}4$, $M\text{=}100$). $T_\f$ is the sequential runtime of $\f$. The results for nnGParareal ($m_{\rm nnGP}\text{=}20$) are not reported as it fails to converge within a 48-hour time budget.}
    \label{tab:swe_speedup}
\end{table}

\section{Discussion and limitations}
\label{sec:discussion}
This study improves the scalability properties, convergence rates, and parallel performance of Parareal and a more recently proposed PinT solver for ODEs and PDEs, nnGParareal~\cite{nngparareal}. By replacing the nnGP with random networks, we decreased the model costs (in learning the discrepancy between the fine and coarse solvers) by several orders of magnitude. 
The reasons behind this are multi-fold. Training of RandNets is cheap due to the availability of 
the closed-form solution for its output (readout) weights, and avoids any expensive hyperparameter optimization. Moreover, it is possible to simultaneously learn and predict the $d$-dimensional correction map instead of $d$ scalar maps (in parallel if the number of processors $N$ is comparable to $d$, or queuing if smaller). The latter ``liberates'' RandNet-Parareal from requiring $d\approx N$,  extending its application to high-dimensional settings, a key/notable improvement with respect to nnGParareal. We tested the proposed algorithm %
on systems of real-world significance, such as the Diffusion-Reaction equation, %
 the SWE, and the Brusselator. %
 solving them on a fine spatial mesh of up to $10^5$ discretization points. %
 These systems and requirements align with those outlined in the benchmark PDE dataset~\cite{takamoto2022pdebench} as necessary prerequisites for using such algorithms in practical scenarios. The strength of RandNet-Parareal is the cheap cost of RandNets, which can be embedded within Parareal 
with virtually no overhead, irrespective of the implementation or solvers, leading to notable speed gains over Parareal (x8.6-21.2 for viscous Burgers', x3-5 for Diffusion-Reaction, x1.3-3.6 for SWE, and x3.4-4.4 for Brusselator). Moreover, training RandNets is easily conducted with established linear algebra routines, and requires no ad-hoc parameter tuning. %

Despite its excellent performance, RandNet-Parareal has limitations common to all Parareal algorithms, as its rate of convergence relies on the accuracy of the coarse solver $\g$. Although neural networks can help mitigate the impact of suboptimal choices of $\g$  (as observed for GPs in (nn)GParareal), if the solver is mismatched for the system — for example, an unstable solver for a stiff ODE — RandNet-Parareal, similar to Parareal and (nn)GParareal, is likely to exhibit non-convergent behavior. %
It would then be of interest to investigate RandNet-Parareal's performance when using customized solvers tailored to specific systems, such as those outlined in Section~\ref{sec:intro} for the shallow water equation and the viscous Burgers' equation, which we defer to future research.

\begin{ack}
\label{sec:ack}
GG is funded by the Warwick Centre of Doctoral Training in Mathematics and Statistics. GG thanks the hospitality of the University of St.~Gallen where part of the results in this paper were obtained.

\end{ack}

\bibliographystyle{abbrvnat}
\bibliography{Bibliography.bib}

\clearpage
\appendix

\section{Pseudocodes}
\label{supp:psudocodes}
This section provides pseudocodes for the implementation of Parareal (Algorithm~\ref{alg:para}), and the training procedure for learning the discrepancy $\f-\g$ via nnGPs in nnGParareal (Algorithm~\ref{alg:nngp}), and RandNets in
RandNet-Parareal (Algorithm~\ref{alg:RandNets}).

\begin{algorithm}
\DontPrintSemicolon
\caption{Parareal (generic)}\label{alg:para}
\KwIn{Initial condition $\bs{u}^0$ at time $t_0$, number of intervals $N$}
\KwOut{Converged initial conditions $\{\boldsymbol{U}_{i}^{K}\}_{i=1}^{N-1}$, with $K$  the number of iterations to convergence}
\;
\textbf{Initialization}\;
Rescale %
    the ODE/PDE system such that each coordinate takes values in $[-1,1]$\;
$L \gets 1$\;
$\bs{U}_0^0 = \bs{u}^0$\;
\For{$i \gets 1$ \textbf{to} $N-1$}{
  $\boldsymbol{U}_{i}^0 \gets \g(\boldsymbol{U}_{i-1}^0)$
}
\;
\For{$k \gets 1$ \textbf{to} $N$}{
  Compute $\f(\boldsymbol{U}_{i-1}^{k-1})$, $i=1,\ldots,N$\ in \textit{parallel}\;
  \For{$i \gets L+1$ \textbf{to} $N-1$}{
      $\boldsymbol{U}_i^k \gets \g(\boldsymbol{U}_{i-1}^{k}) + \fhat(\boldsymbol{U}_{i-1}^{k})$\Comment*[r]{Update the initial conditions}
  }
  \;
  \textbf{Convergence checks}\;
  \For{$i \gets L+1$ \textbf{to} $N-1$}{
    \eIf{$\|\boldsymbol{U}^k_i-\boldsymbol{U}^{k-1}_i\|_{\infty} < \epsilon$}{
        $L \gets L + 1$\Comment*[r]{Update converged interval counter}
    }{
    \textbf{break}\;
    }
  }
    \;
  \If{L == N}{
    \textbf{break}\Comment*[r]{All intervals have converged}
  }
}

\end{algorithm}

\begin{algorithm}
\DontPrintSemicolon
\SetNoFillComment
\caption{nnGP training procedure within nnGParareal}\label{alg:nngp}
\KwIn{Input $\bs{U}_{i-1}^k$, dataset $\mathcal{D}_k$, number of nearest neighbors $m_{\rm nnGP}$, number of random restarts for loss maximization $n_{\rm start}$}
\KwOut{Prediction $\fhat_{\rm nn}(\bs{U}_{i-1}^k)$ of $(\f-\g)(\bs U_{i-1}^k)$}
\;
\textbf{Initialization}\;
\Comment*[l]{Find the $m_{\rm nnGP}$ nns to $\bs{U}_{i-1}^k$, and compute the reduced dataset \eqref{eq:fhat_mnn:U}}
$\mathcal{D}_{i-1,k} \gets \{(\boldsymbol{U}^{(l\mhyphen\textrm{nn})}_{\boldsymbol{U}_{i-1}^k}, \mathbf{Y}^{(l\mhyphen\textrm{nn})}_{\boldsymbol{U}_{i-1}^{k}}),\enspace l=1,\ldots,m_{\rm nnGP} \}\subset \mathcal{D}_k$\;
\;
\tcc{Both loops can be massively parallelized}
\For{$s \gets 1$ \textbf{to} $N$}{
    \tcc{Training}
    \For{$j \gets 1$ \textbf{to} $n_{\rm start}$}{
    \Comment*[l]{Random restarts to avoid local minima when maximizing \eqref{eq:gp_llik}}
    Sample $\boldsymbol{\theta}_j^0$ at random\;
    Maximize \eqref{eq:gp_llik} numerically using $\boldsymbol{\theta}_j^0$ as initial value; obtain $\boldsymbol{\theta}_j^*$\;
    }
  Find $\bs{\theta}^*$ such that 
  \[
  \log p(\widetilde{Y}_{(\cdot,s)}|\widetilde{U}, \boldsymbol{\theta^*}) \geq \log p(\widetilde{Y}_{(\cdot,s)}|\widetilde{U}, \boldsymbol{\theta}_j^*),\quad  j=1,\ldots,n_{\rm start}
  \]\;
  \tcc{Predicting}
  Compute $\mu^{(s)}_{\mathcal{D}_{i-1,k}}(\boldsymbol{U}_{i-1}^{k})$ with (\ref{eq:gp_posterior_m_full}) using $\bs{\theta}^*$\;
}
Set $\fhat_{\rm nn}(\bs{U}_{i-1}^k)\gets(\mu_{\mathcal{D}_{i-1,k}}^{(1)}(\boldsymbol{U}_{i-1}^{k}), \ldots, \mu_{\mathcal{D}_{i-1,k}}^{(d)}(\boldsymbol{U}_{i-1}^{k}))^\top$\;

\end{algorithm}

\begin{algorithm}
\DontPrintSemicolon
\SetNoFillComment
\caption{RandNets training procedure within RandNet-Parareal}\label{alg:RandNets}
\KwIn{Input $\bs{U}_{i-1}^k$, dataset $\mathcal{D}_k$, number of neurons $M$, number of nearest neighbors $m_{\rm RandNet}$}
\KwOut{Prediction $\fhat_{\rm RandNet}(\bs{U}_{i-1}^k)$ of $(\f-\g)(\bs U_{i-1}^k)$}%
\;
\textbf{Initialization}\;
Ensure each ODE/PDE coordinate takes values in $[-1,1]$\;
\Comment*[l]{Find the $m_{\rm RandNet}$ nns to $\bs{U}_{i-1}^k$,  and compute the reduced dataset \eqref{eq:fhat_mnn:U}}
$\mathcal{D}_{i-1,k} \gets \{(\boldsymbol{U}^{(l\mhyphen\textrm{nn})}_{\boldsymbol{U}_{i-1}^k}, \mathbf{Y}^{(l\mhyphen\textrm{nn})}_{\boldsymbol{U}_{i-1}^{k}}),\enspace l=1,\ldots,m_{\rm RandNet} \}\subset \mathcal{D}_k$\;
Sample $A_{w,j} \sim \text{Uniform}(-1,1)$, $w=1,\ldots,M$, $j=1,\ldots,d$\;
Sample $\bs{\zeta}_{w} \sim \text{Uniform}(-1,1)$, $w=1,\ldots,M$\;
Let $\widetilde{X} \in\mathbb{R}^{m\times M}$\;
\;
\textbf{Training}\;
$\widetilde{X}^\top\gets \bs{\sigma}(A\widetilde{U}^\top+\bs{\zeta})$\Comment*[r]{Using broadcasting on $\bs{\zeta}$}
\eIf{${\rm rank}(\widetilde{X}^\top \widetilde{X})==M\leq m$}{
        $\widehat{W}^{\mathcal{D}_{i-1,k}}\gets( \widetilde{X}^\top \widetilde{X} )^{-1}\widetilde{X}^\top \widetilde{Y}$\Comment*[r]{Least-squares estimator}
    }{
    $\widehat{W}^{\mathcal{D}_{i-1,k}}\gets( \widetilde{X}^\top \widetilde{X} )^{\dagger}\widetilde{X}^\top \widetilde{Y}$\;\Comment*[r]{Ridgeless interpolator}
    }
\;
\textbf{Predicting}\;
$\fhat_{\rm RandNet}(\bs{U}_{i-1}^k) \gets (\widehat{W}^{\mathcal{D}_{i-1,k}})^\top \bs{\sigma}(A\bs{U}_{i-1}^k+\bs{\zeta})$\;
\end{algorithm}

\clearpage

\section{Additional details on the nnGParareal correction function}
\label{supp:nngp_full_form}
In this section, we provide more details on the nearest neighbors (nns) Gaussian process modeling, the mathematical expressions of the nnGParareal correction function $\fhat_{\rm nnGPara}$, and the reduced dataset $\mathcal{D}_{i-1,k}$. These are not explicitly presented in the main text as they require additional notation, which we believe does not enrich the explanation. While the description of GPs presented here is for nnGParareal (and the corresponding nnGPs), it immediately generalizes to GParareal by replacing the reduced dataset $\mathcal{D}_{i-1,k}$ with the full dataset $\mathcal{D}_k$. The interested reader can find more details in the original papers,~\cite{nngparareal} and~\cite{pentland2023gparareal}.

Let the set of inputs $\boldsymbol{U}_{i-1}^j \in \mathbb{R}^d$ and outputs $(\f-\g)(\boldsymbol{U}_{i-1}^j)\in \mathbb{R}^d$, $i=1,\ldots, N$, $j=0,\ldots, k-1$, collected by iteration $k$, be denoted by $\mathcal{U}_k$ and $\mathcal{Y}_k$, respectively. Now, define $\mathcal{D}_{i-1,k}$ as the restriction of $\mathcal{D}_k$ to the $m$ nns of $\bs{U}_{i-1}^k$ in $\mathcal{U}_k$, namely
\[
\mathcal{D}_{i-1,k}:= \{(\boldsymbol{U}^{(l\mhyphen\textrm{nn})}_{\boldsymbol{U}_{i-1}^k}, \mathbf{Y}^{(l\mhyphen\textrm{nn})}_{\boldsymbol{U}_{i-1}^{k}}),\enspace l=1,\ldots,m \}\subset \mathcal{D}_k,
\]
where $\mathbf{Y}^{(l\mhyphen\textrm{nn})}_{\boldsymbol{U}_{i-1}^k}=(\f-\g)(\boldsymbol{U}^{(l\mhyphen\textrm{nn})}_{\boldsymbol{U}_{i-1}^k})\in \mathcal{Y}_k$, and $\boldsymbol{U}^{(l\mhyphen\textrm{nn})}_{\boldsymbol{U}_{i-1}^k}$ is the $l$th nn of $\boldsymbol{U}_{i-1}^k$ in $\mathcal{D}_k$, i.e. the $l$th ordered statistics of the set formed out of Euclidean distances $\|\bs U_{i-1}^j-\bs U'\|$ between $\bs U_{i-1}^j$ and any $\bs U'\in \mathcal{U}_k$. That is, there exists $\bs U_1,\ldots, \bs U_{l}=\boldsymbol{U}^{(l\mhyphen\textrm{nn})}_{\boldsymbol{U}_{i-1}^k}\in\mathcal{U}_k$ such that, for any $\bs U'\in\mathcal{U}_k, \bs U'\neq \bs U_r, r=1,\ldots,l$, we have\begin{equation}\label{eq:fhat_mnn:U}
\|\bs U_{i-1}^j-\bs U_1\|\leq \ldots\leq \|\bs U_{i-1}^j-\bs U_{l-1}\|\leq \|\bs U_{i-1}^j-\bs U_l\|\leq \|\bs U_{i-1}^j-\bs U'\|.
\end{equation}
Finally, let ${\widetilde{U}}, {\widetilde{Y}}\in \mathbb{R}^{m\times d}$  be the matrices of input nns and outputs collected in $\mathcal{D}_{i-1,k}$, respectively. 

In nnGParareal, following  the Bayesian framework, a GP prior is placed over the correction function $\f-\g$ for each of the $d$ coordinates as  
\begin{equation*}
(\f-\g)_s \sim GP(\mu_{\rm GP}^{(s)}, \mathcal{K}_{\rm GP}), \enspace s=1,\ldots, d,
\end{equation*}
where $\mu_{\rm GP}^{(s)}: \R^d \rightarrow{} \R$ is the prior mean function, taken to be zero for all $s=1,\ldots, d$, and $\mathcal{K}_{\rm GP}: \R^d \times \R^d \rightarrow \R$ is the exponential prior variance kernel function 
\[\mathcal{K}_{\rm GP}(\boldsymbol{U},\boldsymbol{U}') = \sigma_{\rm o}^2 \exp(-\|\boldsymbol{U}-\boldsymbol{U}'\|^2 / \sigma_{\rm in}^2),
\]
with $\sigma_{\rm in}^2$ and $\sigma_{\rm o}^2$ denoting the input and output length scales, respectively. Differently from the prior mean, the prior variance is the same across the $d$ components. Then, each nnGParareal prediction $\fhat_{\rm nnGPara}^{(s)}(\boldsymbol{U}_{i-1}^k)\in \mathbb{R}$, $s=1,\ldots, d$, is obtained from the GP posterior mean $\mu_{\mathcal{D}_{i-1,k}}^{(s)}(\boldsymbol{U}_{i-1}^k) \in \mathbb{R}$, computed on the reduced dataset $\mathcal{D}_{i-1,k}$,  given by
\begin{equation}
    \fhat_{\rm nnGPara}^{(s)}(\boldsymbol{U}_{i-1}^k) = \mu^{(s)}_{\mathcal{D}_{i-1,k}}(\boldsymbol{U}_{i-1}^k):= \mathcal{K}(\widetilde{U}, \boldsymbol{U}_{i-1}^k)^\top (\mathcal{K}(\widetilde{U},\widetilde{U}) + \sigma_{\rm reg}^2 \mathbb{I}_{m})^{-1} \widetilde{Y}_{(\cdot,s)}, \label{eq:gp_posterior_m_full}
\end{equation}
where $\mathcal{K}(\widetilde{U}, \boldsymbol{U}_{i-1}^k)\in \mathbb{R}^{m}$ is a vector of covariances between every input collected in $\widetilde{U}$ and  $\boldsymbol{U}_{i-1}^k$ defined as $(\mathcal{K}(\widetilde{U},\boldsymbol{U}_{i-1}^k))_{r}=\mathcal{K}_{\rm GP}({(\widetilde{U}_{{(r,\cdot)}}})^\top, \boldsymbol{U}_{i-1}^k )$, $r=1,\ldots, m$%
, and %
$\mathcal{K}(\widetilde{U},\widetilde{U})\in \mathbb{R}^{m\times m}$ is the covariance matrix, with $(\mathcal{K}(\widetilde{U}, \widetilde{U}))_{q,r} = \mathcal{K}_{\rm GP}({(\widetilde{U}_{{(q,\cdot)}}}) ^\top, {(\widetilde{U}_{{(r,\cdot)}}})^\top )$, $r,q=1,\ldots, m$. %
Here, $\sigma^2_{\rm reg}$ denotes a regularization term, also known as nugget, jitter, or regularization strength, which is added to improve the numerical stability when computing the inverse matrix, see~\cite{nngparareal} for further details. The hyperparameters $\boldsymbol{\theta}:=(\sigma_{\rm in}^2, \sigma_{\rm o}^2, \sigma_{\rm reg}^2)$ entering into the posterior mean and prediction \eqref{eq:gp_posterior_m_full}
control the performance of the GP,  and are optimized by numerically maximizing the marginal log-likelihood:
\begin{equation}
   \log p(\widetilde{Y}_{(\cdot,s)}|\widetilde{U}, \boldsymbol{\theta}) \propto - {\widetilde{Y}_{(\cdot,s)}}^\top(\mathcal{K}(\widetilde{U},\widetilde{U})+\sigma_{\rm reg}^2 \mathbb{I}_{m})^{-1}\widetilde{Y}_{(\cdot,s)} - \log {\rm det}(\mathcal{K}(\widetilde{U},\widetilde{U})),
    \label{eq:gp_llik}
\end{equation}
where $\mathcal{K}(\cdot, \cdot)$ depends on $\boldsymbol{\theta}$ through the kernel $\mathcal{K}_{\rm GP}$, and $\det(A)$ denotes the determinant of a square matrix $A$. For a thorough treatment of Gaussian processes, including derivation of the likelihood  and of the posterior distribution (which is Gaussian with mean as in~\eqref{eq:gp_posterior_m_full}, see~\cite{williams2006gaussian}.

\section{Computational complexity analysis}
\label{supp:comp_complx}
Consider the $d$-dimensional initial value problem \eqref{eq:ode} for some (O/P)DE. Let $N$ be the number of subintervals (data points) at each $k$th iteration of the PinT algorithm. For any $k$th iteration of the scheme, a total of $Nk$ data points, each $d$-dimensional, are available. Here, we provide the computational cost of RandNet-Parareal, and compare it to that of nnGParareal, the state-of-the-art Parareal algorithm proposed in~\cite{nngparareal}. Both RandNet-Parareal and nnGParareal use only the reduced data set of $m$ nns to a given point to construct its image-prediction via \eqref{eq:update_rule_generic}.  Note that the  $m$ nns (in Euclidean distance) to some point $\bs{U}\in \mathbb{R}^d$ among $Nk$ available points are found at a cost which is at most linear in the sample size, that is $O(mNk)$ (for moderate dimensions $d$, one can get an improved cost $O(m\log(Nk))$, logarithmic in the sample size) \cite{nngparareal}. Since our goal is to compare the computational complexities of nnGParareal and RandNet-Parareal as a function of $d$, we consider the worst-case complexity of the nns search.

Given an input $\bs{U}_{i-1}^k \in \mathbb{R}^d$, $i=1,\ldots, N$ at iteration $k$, the computational model cost of a prediction $\bs{U}_{i-1}^k$ 
produced by all $d$ models of $m_{\rm nnGP}$-nnGPs at iteration $k$ 
via the predictor-corrector rule \eqref{eq:update_rule_generic} with 
nnGParareal correction \eqref{eq:gp_posterior_m_full} and $m_{\rm nnGP}$ nns is given in \cite{nngparareal} as
\begin{align*}
   T_{\rm nnGP}(k) &\leq  C_{\rm{nnGP}}  Nk( n_{\rm start}n_{\rm reg}\frac{d}{N} \vee 1)  \times\\&(\underbrace{m_{\rm nnGP}d}_{B:=\mathcal{K}(U, \bs{U}_{i-1}^{k-1})^\top}+ \underbrace{m_{\rm nnGP}^2d}_{C:=\mathcal{K}(U,U)}+ \underbrace{m_{\rm nnGP}^3}_{D:=(B+\sigma^2_{\rm reg}\mathbb{I}_{m_{\rm nnGP}})^{-1}} + \underbrace{m_{\rm nnGP}^2}_\text{$B \cdot D$}+\underbrace{m_{\rm nnGP}d}_{BD \cdot Y}+\underbrace{m_{\rm nnGP}Nk}_\text{nearest neighbors})\\&=  C_{\rm{nnGP}}Nk (n_{\rm reg} n_{\rm start} \frac{d}{N} \vee 1)   ({m_{\rm nnGP}^3} +{m_{\rm nnGP}^2}+d(m^2_{\rm nnGP}+2m_{\rm nnGP})+{m_{\rm nnGP}Nk}),
\end{align*}
with $C_{\rm{nnGP}}$ being some constant  that in general {\it does depend} on $k$, $m_{\rm nnGP}$, and $d$. Also, $n_{\rm reg}$ and $n_{\rm start}$ correspond to the number of random restarts and the number of explored values of the regularization penalty in the kernel regression (associated to the hyperparameter optimization (see \cite[Section 4.5]{nngparareal}), respectively. Furthermore, $\vee$ is the maximum operator, and the factor $(n_{\rm start}n_{\rm reg}d/N \vee 1) \geq 1$ follows from the fact that  $d$ independent nnGPs and hyperparameter optimization are parallelized over the $N$ cores.

In RandNet-Parareal,  the correction term $\fhat_{\rm RandNet\text{-}Para}$ is modeled by the random weights neural network and evaluated as \eqref{eq:randnet}. Again, only 
  $m_{\rm RandNet}$  nns (in Euclidean distance) to $\bs{U}_{i-1}^k$ are used to construct the prediction, leading to the following  computational model cost at iteration $k$:
\begin{align*}
T_{\rm RandNet}(k) \leq  C_{\rm{RandNet}}Nk&\frac{1}{N}(\underbrace{Mdm_{\rm RandNet}}_{{X}:=\boldsymbol{\sigma}(A \cdot U + \boldsymbol{\zeta})} + \underbrace{M^2m_{\rm RandNet}}_{\Sigma:=X\cdot X^\top}+ \underbrace{M r^2}_{\Sigma^{\dagger}} \\&+ \underbrace{Mm_{\rm RandNet}d}_{\Sigma^{\dagger} \cdot X}+\underbrace{M^2 d}_{W:=\Sigma^{\dagger} X\cdot Y}+\underbrace{Md m_{\rm RandNet}}_{W^\top \cdot X}+\underbrace{m_{\rm RandNet}Nk}_\text{nearest neighbors})\\=C_{\rm{RandNet}}k(&M r^2 +  {M^2m_{\rm RandNet}}+ d(M^2+3M m_{\rm RandNet})+{m_{\rm RandNet}Nk}),
\end{align*}
where $M$ is the number of hidden neurons, $r$ is the rank of the covariance of activated neurons $\Sigma$ (mind that the pseudoinverse of $\Sigma$ would contribute cubically in $m$ only if $\Sigma$ is of full rank numerically, which is not observed empirically) and $C_{\rm{RandNet}} $ is 
a constant  {\it independent} on $N$, $k$, $M$, $d$, $m_{\rm RandNet}$. The factor $1/N$ in the first inequality corresponds to  parallelization over $N$ processors.

\begin{figure}
    \centering\vspace{-0.7cm}
    \includegraphics[width=1\linewidth,trim={0 0cm 0  0cm},clip]{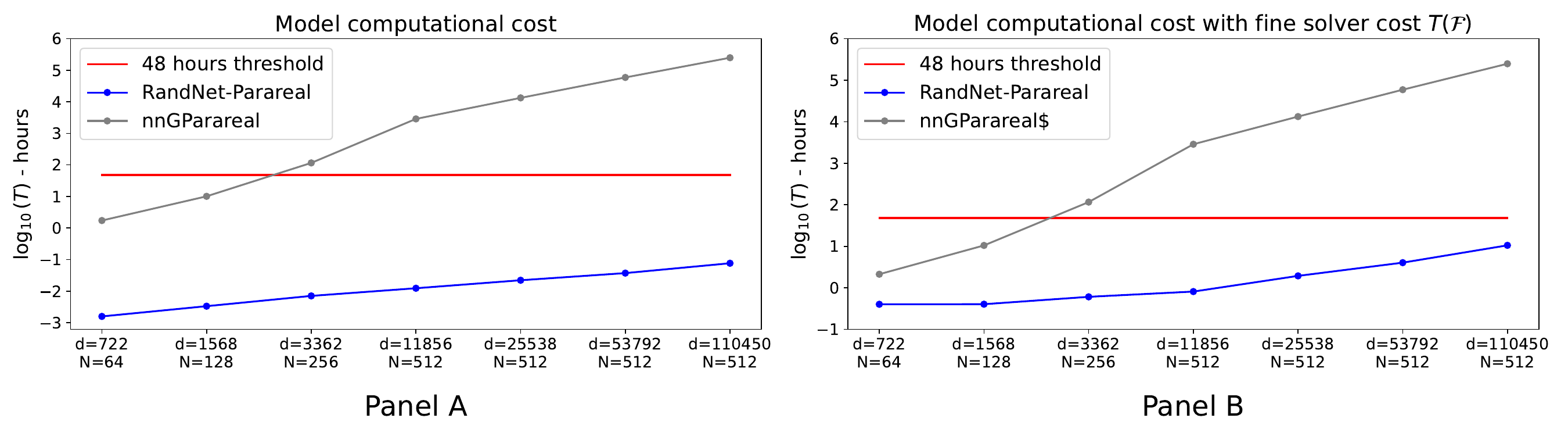}
    \caption{Theoretical {\it model} cost (panel A) and theoretical {\it total} cost (panel B), as functions of the dimension $d$ (and the corresponding $N$). The results are reported in terms of $\log_{10}({\rm hours})$.}
    \label{fig:comp_complx}
\end{figure}

We note the following differences in costs between these two algorithms according to realistic situations:
\begin{itemize}
    \item  $d\gg N$ in most relevant applications, especially for PDEs. Hence, $(n_{\rm start}n_{\rm reg}d/N \vee 1) \gg 1$, limiting the benefits from parallelization for nnGParareal. In the considered experiments, we had access to a maximum of approximately $N=500$ processors, while we considered up to $d \approx 10^5$. It is easy to see that $T_{\rm nnGP}$ is quadratic in dimension $d$, while $T_{\rm RandNet}$ is only linear. This difference is mainly due to the factor $(n_{\rm start}n_{\rm reg}d/N \vee 1) \geq 1$ in $T_{\rm nnGP}$ as opposed to $1/N$ in $T_{\rm RandNet}$.
    \item Although $M > m_{\rm nnGP}$, $M=100$ is sufficient for consistent performance across a range of systems, as shown in our numerical experiments.
    \item nnGParareal incurs additional cost due to hyperparameter optimization~\cite{nngparareal}, necessary for tuning the kernel input and output scales and the regularization strength for each of the $d$ dimensions, which is performed by maximizing the loglikelihood. First, to explore the parameter space and allow for multiple starting points given the nonconvex optimization problems, both $n_{\rm reg}$ and $n_{\rm start}$ should be set large. Second, each loglikelihood maximization conducted per dimension of the system requires a large number of iterations performed sequentially.  Hence, $C_{\rm{nnGP}}\gg C_{\rm{RandNet}}$, with $C_{\rm{nnGP}}$ depending, in general, on $k$, $m_{\rm nnGP}$, $d$, as opposed to $C_{\rm{RandNet}}$. Indeed, RandNet requires no tuning (a significant advantage with respect to GPs and nnGPs, making it more user-friendly), neither for the distribution of the random weights nor for the regularization parameter $\lambda$, due to the use of the ridgeless estimator. Empirically, we observed $C_{\rm nnGP}/C_{\rm RandNet}$ to be up to $ 1000$ (this can be seen in Figure \ref{fig:diff_react_speedup}).
    \item We emphasize that since the ReLu function is chosen as activation in RandNet, the matrix $X$ of activated neurons is sparse with sparsity degree $\gamma$. Hence, the computational complexity $T_{\rm RandNet}$ could be further improved, as the  computational complexity of sparse operations is proportional to the number of nonzero elements in the matrix. We intentionally left these arguments out of the complexity analysis, since we do not use sparse operations in our code implementation.
    \item The upper bound of $T_{\rm RandNet}$ could potentially be improved further, as additional parallelization may occur during standard matrix operations, depending on the specific computing environment.
\end{itemize}
Figure \ref{fig:comp_complx} illustrates the theoretical {\it model} costs $T_{\rm nnGP}$ and $T_{\rm RandNet}$ (Panel A) and theoretical {\it total} costs obtained by adding the coarse and fine solver costs (Panel B), as functions of the dimension $d$ (and the corresponding $N$). The results are reported in terms of $\log_{10}({\rm hours})$. To calibrate the constants in both complexity bounds, we used the total empirical computational cost in Figure \ref{fig:diff_react_speedup}, together with its breakdown described in Table \ref{tab:diffreact_speedup}. Panel A shows that RandNet-Parareal displays significant improvement in scalability with respect to the state-of-the-art Parareal algorithm nnGParareal, while Panel B demonstrates that whenever the cost of the fine solver is added, our results are in full coherence with the empirical results.

\section{Robustness study}
\label{supp:robustness}
\begin{figure}[h!]
     \centering     \includegraphics[width=1\linewidth]{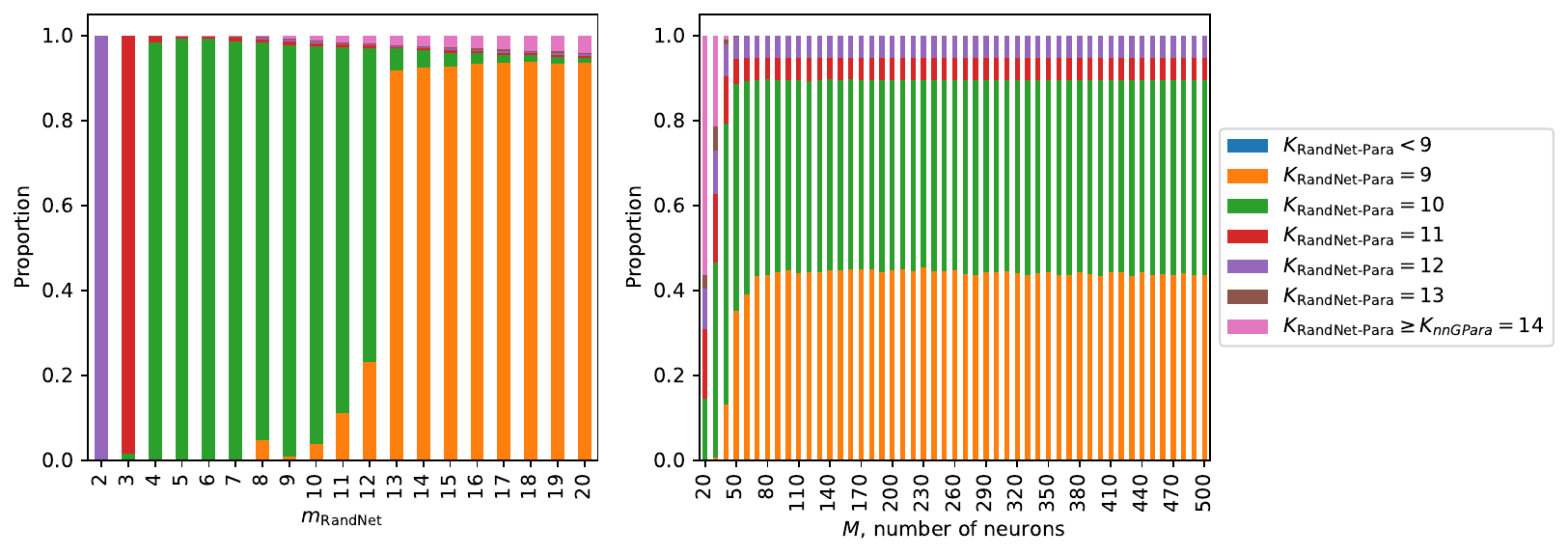}
     \caption{Histogram of the iterations to convergence $K_{\rm RandNet\text{-}Para}$ of RandNet-Parareal for $d=128$ for Burgers' equation. We sample the network weights $A$, $\bs{\zeta}$ 100 times. For each set of weights, we run RandNet-Parareal for $m_{\rm RandNet} \in \{2, 3,\ldots, 20\}$ and $M \in \{20, 30, 40, \ldots, 500\}$. The left and right panels show the aggregated histograms of $K_{\rm RandNet\text{-}Para}$ versus $m_{\rm RandNet}$ and $M$, respectively.}
     \label{fig:Burgers_perf_across_m}
 \end{figure}

\begin{figure}[h!]
     \centering     \includegraphics[width=1\linewidth]{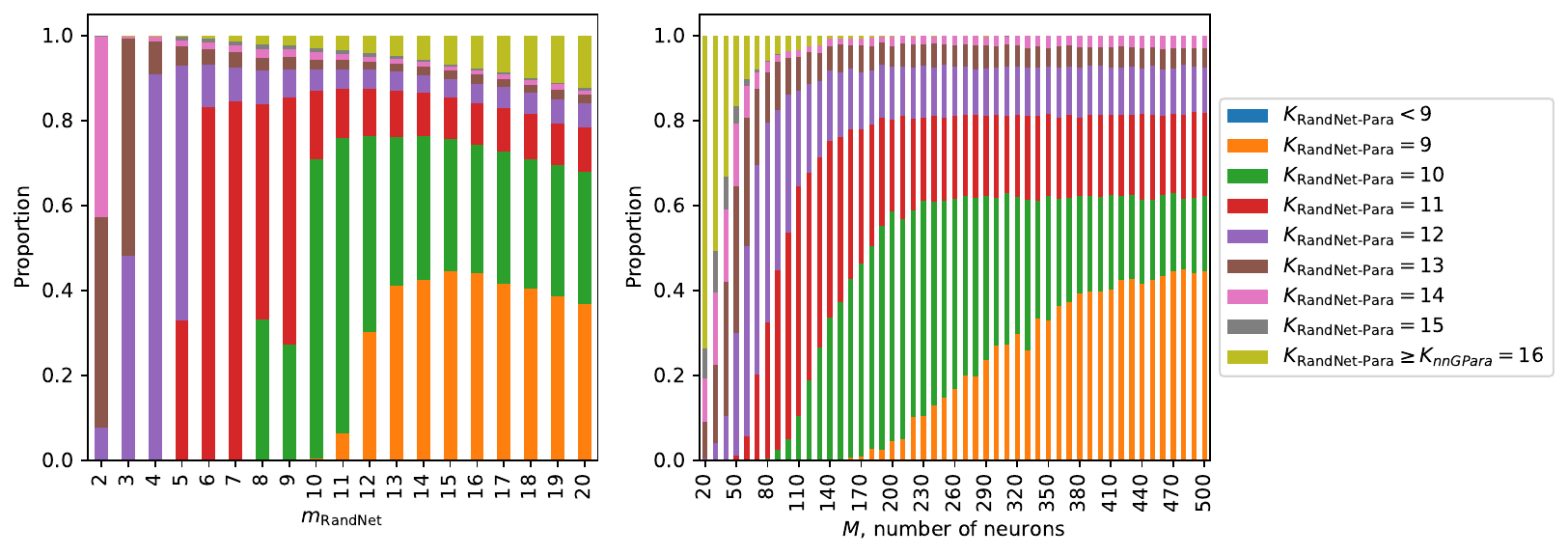}
     \caption{Histogram of the iterations to convergence $K_{\rm RandNet\text{-}Para}$ of RandNet-Parareal for $d=722$ for Diffusion-Reaction equation. We sample the network weights $A$, $\bs{\zeta}$ 100 times. For each set of weights, we run RandNet-Parareal for $m_{\rm RandNet} \in \{2, 3,\ldots, 20\}$ and $M \in \{20, 30, 40, \ldots, 500\}$. The left and right panels show the aggregated histograms of $K_{\rm RandNet\text{-}Para}$ versus $m_{\rm RandNet}$ and $M$, respectively.}
     \label{fig:DiffReact_perf_across_m}
 \end{figure}

In this section, we study the robustness of RandNet-Parareal to changes in the number of nns $m_{\rm RandNet}$, the number of neurons $M$, and the randomly  sampled values of neural network weights $A$, $\bs{\zeta}$. Our empirical findings (for two of the three considered PDEs) demonstrate that the iterations $K_{\rm RandNet\text{-}Para}$ to convergence for RandNet-Parareal remain largely consistent despite variations in these factors. This ensures robust performance across a broad spectrum of parameter values, reducing users' need for extensive tuning. %
For computational tractability, we limit the robustness analysis to relatively small systems, such as  Burgers' equation with $d=128$, and 
the Diffusion-Reaction equation with $d=722$, conducting 100 weight samplings for each system. For every set of weights, we iterate RandNet-Parareal across $m_{\rm RandNet}$ values ranging from 2 to 20, and $M$ values ranging from 20 to 500 in increments of 10. The proportions of iterations needed to converge across 100 runs for different values of $m_{\rm RandNet}$ and $M$ for the Burger's and diffusion-Reaction equations are reported in Figures~\ref{fig:Burgers_perf_across_m} and~\ref{fig:DiffReact_perf_across_m}, respectively. 
Although we observe some minor differences between the two systems, the main trend is clear: as long as reasonable values of $m_{\rm RandNet}$ and $M$ are chosen, the iterations to convergence for RandNet-Parareal vary at most by a few units  when changing the values of
$m_{\rm RandNet}$, $M$ or a particular sampling seed of weights. %
Nevertheless, larger $M$ %
might improve the performance, since, in this case, RandNets operate in the interpolation regime, as discussed in Section \ref{sec:elm}.

\section{Additional numerical experiments: 2D and 3D Brusselator PDE}
\label{supp:brusselator}

\begin{figure}[h!]
    \centering
    \includegraphics[width=1\linewidth, trim={0 0cm 0  0cm},clip]{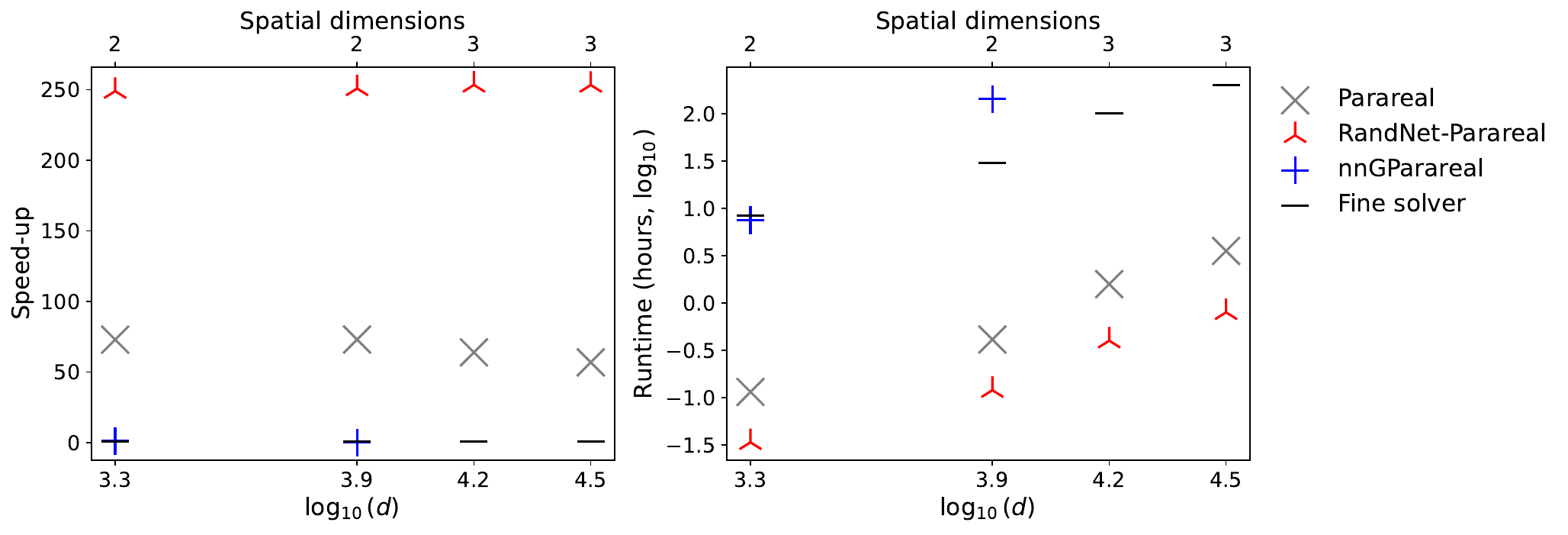}
    \caption{Scalability study for the 2 and 3 spatial dimensional Brusselator PDE. We used $N=512$ and 48 hours runtime budget. nnGParareal for $\log_{10}(d)=3.9$ is estimated, as the algorithm does not converge within the 48 hours runtime budget.}
    \label{fig:brusselator}
\end{figure}

Here, we carry out an additional %
scalability study for the 2 and 3 spatial dimensional Brusselator PDE. This model is a two-component reaction system that exhibits complex behavior, including oscillations, spatial patterns, and chaos. It is described by 
\[ \partial_t u = D_0 \nabla^2 u + a - (1 + b) u + vu^2,\] 
and 
\[\partial_t v = D_1 \nabla^2 v + bu - vu^2.\]
In chemistry, the components $u,v$ refer to the concentration of two substances, whereas the constants $D_0, D_1$ are the respective diffusivity of each component, indicating the rate at which the substances spread out in space. Moreover, the parameters $a$ and $b$ are related to reaction rates. In our experiments, we used $D_0=0.1$, $D_1=0.1 D_0$, $a=1$, and $b=3$. We take $t \in [0, 35]$, $(u,v)\in (-1,1)^2\times (-1,1)^2$ for the 2D Brusselator, and $(u,v)\in (-1,1)^3\times (-1,1)^3$ for the three spatial dimension case. We initialize the $u$ values at time $t=0$ by setting them equal to $a$, and the $v$ values by taking them normally distributed over the spatial grid. Further details regarding the number of spatial discretizations, the number of intervals $N$ and the order of the solvers $\f$ and $\g$ is given in Table \ref{tab:brus_setup}.
\begin{table}[t]
    \centering
    \footnotesize
    \caption{Simulation setup for the 2D and 3D Brusselator}
    \begin{tabular}{cccccccc}
    \toprule
       Domain & $N_u = N_v$ &  $d$&  $\g$ & $\g_{\Delta t}$ & $\f$ & $\f_{\Delta t}$ & $N$  \\
       \midrule
          $(u,v,t) \in (-1,1)^2\times (-1,1)^2\times[0,35]$ &32 &2048 &RK1 &0.034 &RK4 &$1e^{-7}$ & 512\\
          $(u,v,t) \in (-1,1)^2\times (-1,1)^2\times[0,35]$ &64 &8192 &RK1 &0.033 &RK4 &$1e^{-7}$ &512\\
          $(u,v,t) \in (-1,1)^3\times (-1,1)^3\times[0,35]$ &20 &16000 &RK1 &0.052 &RK4 &$1e^{-7}$ &512\\
          $(u,v,t) \in (-1,1)^3\times (-1,1)^3\times[0,35]$ &25 &31250 &RK1 &0.057 &RK4 &$1e^{-7}$ &512\\
    \bottomrule
    \end{tabular}\\
    \caption*{%
    $N_u$ and $N_v$ are the number of spatial discretization points for $u$ and $v$ along each spatial dimension, yielding a $d=2N_x^2$- or $d=3N_x^3$-dimensional ODE, depending on the considered system. $\g$ and $\f$ denote the coarse and fine solvers, respectively, while the $_{\Delta t}$ subscript refers to the timestep. The number of nns used for $\mathcal{D}_{i-1,k}$ in nnGParareal and RandNet-Parareal  are  
    $m_{\rm nnGP}=20$ and $m_{\rm RandNet}=4$,  respectively. $N$ is the total number of intervals.}
    \label{tab:brus_setup}
\end{table}
Figure \ref{fig:brusselator} highlights the strong scaling advantages of RandNet-Parareal compared to nnGParareal, setting $N=512$ and restricting the runtime budget to a maximum of 48 hours, as done in the other test cases. 

\section{Accuracy and runtimes across models and algorithms}
\label{supp:accuracy}
In Table \ref{tab:accuracy} below, we report the accuracies and runtimes (shown in parentheses) for RandNet-Parareal, Parareal, and nnGParareal. The accuracy is measured with maximum absolute error (mean across intervals) with respect to the true solution obtained by running $\f$ sequentially. Interestingly, all accuracies are far below the pre-defined accuracy level $\epsilon$, with RandNet-Parareal achieving the lowest one in all but one experiment, with much smaller runtimes across all case studies. 
\begin{table}[h!]
\footnotesize
    \caption{Accuracy and computational cost of the three considered algorithms}
        \label{tab:accuracy}
\centering
\begin{tabular}{@{}lccc@{}}
\toprule
\textbf{PDE} & \textbf{RandNet-Parareal} & \textbf{Parareal} & \textbf{nnGParareal} \\ \midrule
Burgers' $d = 128$ & $1.06e^{-8}$ (1h 2m) & $1.85e^{-8}$ (8h 54m) & $1.32e^{-7}$ (1h 39m) \\ 
Diffusion-Reaction $d = 7.2e^{2}$ & $3.56e^{-8}$ (23m) & $1.83e^{-8}$ (1h 40m) & $5.71e^{-7}$ (1h 11m) \\ 
Diffusion-Reaction $d = 3.3e^{3}$ & $8.56e^{-10}$ (33m) & $2.45e^{-8}$ (7h 52m) & not converged \\ 
Diffusion-Reaction $d = 2.5e^{4}$ & $8.09e^{-11}$ (1h 57m) & $7.43e^{-9}$ (9h 50m) & not converged \\ 
SWE $d = 3.1e^{4}$ & $6.75e^{-8}$ (4h 9m) & $5.15e^{-8}$ (15h 43m) & not converged \\ 
SWE $d = 6.1e^{4}$ & $8.54e^{-9}$ (12h 34m) & $2.84e^{-8}$ (19h 30m) & not converged \\ 
Brusselator 2D $d=2e^3$ & $2.09e^{-8} $ (2m) & $3.16e^{-8}$ (7m) & $3.38e^{-7}$ (7h 31m)\\
\bottomrule
\end{tabular}
\caption*{Accuracy and computational cost comparison of RandNet-Parareal, Parareal, and nnGParareal for different PDEs, with  runtimes reported in parentheses. The accuracy is measured as maximum absolute error (mean across intervals) with respect to $\f$ run sequentially.}
\end{table}

\section{Simulation setups}
\label{supp:sim_setup}
This section summarizes the simulation setups used for producing the results discussed in Section~\ref{sec:experiments} in the main text. The tables below report the space and time domain of the considered PDEs, the number of spatial discretization points $N_x$ (and $N_y$, in case of two-dimensional spatial systems),  the numerical solvers used for $\g$ and $\f$, their corresponding numbers of time steps per interval, the number of intervals $N$, and the number of nns used for nnGParareal ($m_{\rm nnGP}$) and RandNet-Parareal ($m_{\rm RandNet}$). In particular, Table~\ref{tab:burg_summ_setup} refers to the viscous Burgers' equation,  Table \ref{tab:diffreact_summ_setup} to the Diffusion-Reaction equation, and  Table \ref{tab:swe_summ_setup} to the shallow water equations (SWEs).

\begin{table}[t]
    \centering
    \footnotesize
    \caption{Simulation setup for the viscous Burgers' equation}
    \begin{tabular}{cccccccccc}
    \toprule
       Domain & $N_x$ & $d$&  $\g$ & $N_\g$ & $\f$ & $N_\f$ & $N$ & $m_{\rm nnGP}$ & $m_{\rm RandNet}$  \\
       \midrule
         $(x,t)\in [-1,1]\times[0,5.9]$ & 128 &128& RK1 & 4 & RK8 & $4e^4$ & 128 & 18 & 3\\
         $(x,t)\in [-1,1]\times[0,5.9]$ & 1128 &1128& RK1 & 293 & RK8 & $6e^5$ & 128 & 18 & 3\\
    \bottomrule
    \end{tabular}\\
    \caption*{%
    $N_x$ is the number of space discretizations, the same as $d$ here. $\g$ and $\f$ denote the chosen  coarse and fine solvers, with  corresponding time discretization steps per interval $N_\g$ and $N_\f$,  respectively. Here $N$ is the number of intervals, while $m_{\rm nnGP}$ and $m_{\rm RandNet}$ are the numbers of nns used to create $\mathcal{D}_{i-1,k}$ for nnGParareal and RandNet-Parareal, respectively.}
    \label{tab:burg_summ_setup}
\end{table}

\begin{table}[t]
    \centering
    \footnotesize
    \caption{Simulation setup for the Diffusion-Reaction equation}
    \begin{tabular}{ccccccccc}%
    \toprule
       Domain & $N_x$ & $N_y$ & $d$ & $\g$ & $N_\g$ & $\f$ & $N_\f$ & $N$ \\%
       \midrule
         $(x,y,t)\in [-1,1]^2\times[0,20]$ & 19 & 19 & 722 & RK1 & 1 & RK4 & NA & 64 \\%
         $(x,y,t)\in [-1,1]^2\times[0,20]$ & 28 & 28 & 1568 & RK1 & 1 & RK4 & NA & 128  \\%
         $(x,y,t)\in [-1,1]^2\times[0,20]$ & 41 & 41 & 3362 & RK1 & 1 & RK4 & NA & 256  \\%
         $(x,y,t)\in [-1,1]^2\times[0,20]$ & 77 & 77 & 11858 & RK4 & 1 & RK8 & NA & 512  \\%
         $(x,y,t)\in [-1,1]^2\times[0,20]$ & 113 & 113 & 25538 & RK4 & 2 & RK8 & NA & 512  \\%
         $(x,y,t)\in [-1,1]^2\times[0,20]$ & 164 & 164 & 53792 & RK4 & 4 & RK8 & NA & 512  \\%
         $(x,y,t)\in [-1,1]^2\times[0,20]$ & 235 & 235 & 110450 & RK4 & 8 & RK8 & NA & 512  \\%
         \bottomrule
    \end{tabular}\smallskip\\
    \caption*{%
    $N_x$ and $N_y$ are the number of spatial discretization points for $x$ and $y$, respectively,  yielding a $d=2N_xN_y$-dimensional ODE.  $\g$ and $\f$ denote the coarse and fine solvers, respectively. The number of nns used for $\mathcal{D}_{i-1,k}$ in nnGParareal and RandNet-Parareal  are  
    $m_{\rm nnGP}=20$ and $m_{\rm RandNet}=3$,  respectively. $N_\g$ is the time discretization steps of $\g$ per interval. $N_\f= {\rm NA}$ since  $\f$'s step size is  chosen by \textit{scipy} Runge-Kutta method~\cite{virtanen2020scipy}. }
    \label{tab:diffreact_summ_setup}
\end{table}

\subsection{Simulation setup for the SWEs}
\label{app:SWE}
Here, we give more details on the radial dam break simulation of Section~\ref{sec:swe}. Our domain consists of a  rectangular box defined as $(x,y) \in [-5,5]\times [0,5]$, which we evolve temporally over $t \in [0, 20]$. Following~\cite{takamoto2022pdebench}, as an initial condition, we place a Gaussian-shaped column of water centered at $(x,y)=(-2.5, 1.5)$, with covariance matrix $\Sigma= \left(
\begin{array}{cc}
0.25&0 \\
0&0.25 \\
\end{array}
\right)$. We use Neumann boundary conditions, and evolve the system using $N=235$ intervals over four increasingly finer spatial meshes, as described in Table \ref{tab:swe_summ_setup}. We used the \textit{ParareaML}~\cite{PararealML} Python package to implement the SWEs and corresponding numerical solvers. %

\begin{table}[ht]
    \centering
    \footnotesize
    \caption{Simulation setup for the SWEs}
    \begin{tabular}{ccccccccc}%
    \toprule
       Domain & $N_x$ & $N_y$ & $d$ & $\g$ & $N_\g$ & $\f$ & $N_\f$ & $N$ \\%
       \midrule
         $(x,y,t)\in [-5,5]\times[0,5]\times[0,20]$ & 101 & 51 & 15453 & RK1 & 7 & RK4 & $1e^5$ & 235\\%
         $(x,y,t)\in [-5,5]\times[0,5]\times[0,20]$ & 144 & 72 & 31104 & RK1 & 8 & RK4 & $2e^5$ & 235 \\%
         $(x,y,t)\in [-5,5]\times[0,5]\times[0,20]$ & 201 & 101 & 60903 & RK1 & 14 & RK4 & $4e^5$ & 235 \\%
         $(x,y,t)\in [-5,5]\times[0,5]\times[0,20]$ & 264 & 133 & 105336 & RK1 & 24 & RK4 & $5e^5$ & 235\\%
         \bottomrule
    \end{tabular}\smallskip\\
    \caption*{$N_x$ and $N_y$ are the number of spatial discretization points for $x$ and $y$, respectively, leading to an ODE of  dimension $d=3N_x N_y$. $\g$ and $\f$ denote the chosen numerical coarse and fine solvers, respectively, with $N_\g$ and $N_\f$ being their corresponding time discretization steps per interval. In all cases, we set the number of nns used to create $\mathcal{D}_{i-1,k}$ to $m_{\rm nnGP}=20$ 
 for nnGParareal, and $m_{\rm RandNet}=3$ for RandNet-Parareal.}
    \label{tab:swe_summ_setup}
\end{table}

\clearpage 
\section{Illustration of some PDE solutions}
\label{supp:illustrations}

 \begin{figure}[ht]
     \centering     \includegraphics[width=0.5\linewidth]{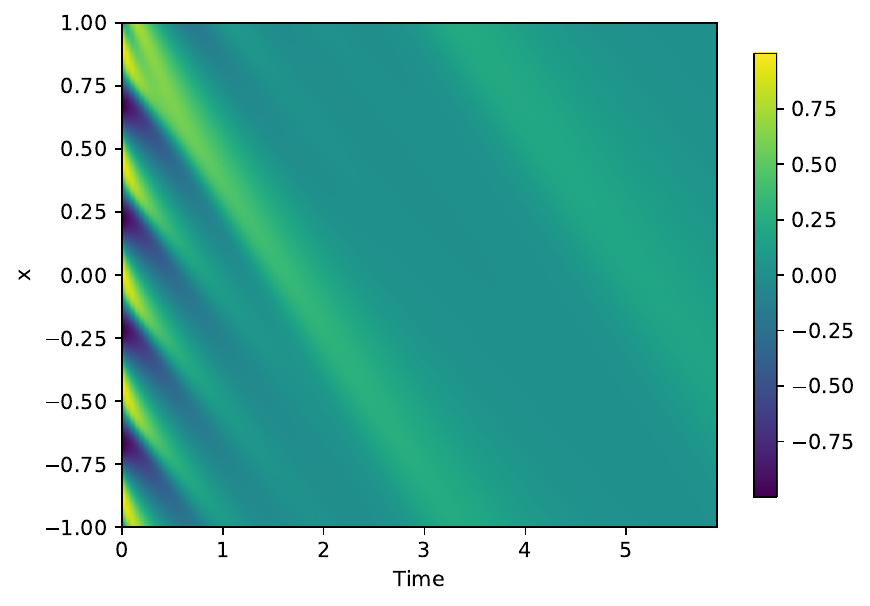}
     \caption{Numerical solution of viscous Burgers' equation over $(x,t) \in [-1,1]\times[0,5.9]$ with $d=1128$ and initial conditions and additional settings as described in Section~\ref{sec:burgers}.}
     \label{fig:burgers_system_evolution}
 \end{figure}
 
\begin{figure}[ht]
     \centering     \includegraphics[width=1\linewidth]{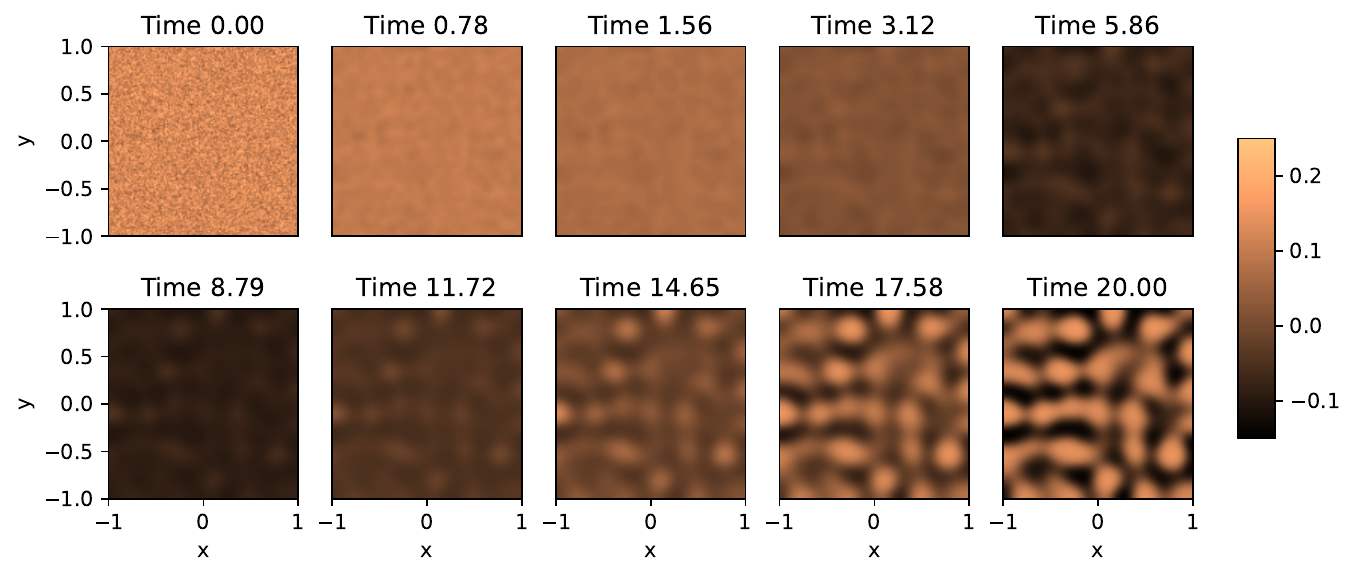}
     \caption{Numerical solution of the Diffusion-Reaction equation over $(x,y) \in [-1,1]^2$ with $N_x=N_y=235$ for a range of system times $t$. Only the activator $u(t,x,y)$ is plotted. The initial conditions and additional settings are as described in Section~\ref{sec:diffreact}.}
     \label{fig:diffreact_system_evolution}
 \end{figure}
 
\section{Additional simulation results for the Diffusion-Reaction equation}
\label{supp:DiffReactTable}
Here, we complement the results of the speed-ups and wallclock times reported in Figure~\ref{fig:diff_react_speedup} in the main text, with a detailed breakdown of the number of iterations to convergence, the runtimes of the coarse and fine solvers, the overall cost of training the model (up to convergence), and the total runtime, reported in Table \ref{tab:diffreact_speedup}.

\begin{table}[t]
    \footnotesize
    \centering
    \caption{Speed-up analysis for the Diffusion-Reaction equation}
   
    \begin{tabular}{L{2.3cm}C{0.3cm}cC{1.1cm}C{1.3cm}C{1.3cm}c}
    \addlinespace
    \multicolumn{7}{c}{$d=722$, $N=64$}\\
    \toprule
    Algorithm & $K$ & $NT_{\g}$ & $T_{\f}$ & $T_{\rm model}$ & $T_{\rm alg}$ & $S_{\rm alg}$\\
    \midrule
    Fine & $-$ & $-$ & $-$ & $-$ & 2h 2m & 1\\
    Parareal & 53 & 0s & 2m & 0s & 1h 40m & 1.22\\
    nnGParareal & 16 & 0s & 2m & 40m & 1h 11m & 1.72\\
    RandNet-Parareal & 12 & 0s & 2m & 1s & 23m & \textbf{5.36}\\
    \bottomrule
    \addlinespace
    \multicolumn{7}{c}{$d=1568$, $N=128$}\\
    \toprule
    Algorithm & $K$ & $NT_{\g}$ & $T_{\f}$ & $T_{\rm model}$ & $T_{\rm alg}$ & $S_{\rm alg}$\\
    \midrule
    Fine & $-$ & $-$ & $-$ & $-$ & 4h 21m & 1\\
    Parareal & 93 & 0s & 2m & 0s & 3h 3m & 1.42\\
    nnGParareal & 20 & 0s & 2m & 10h 32m & 11h 12m & 0.39\\
    RandNet-Parareal & 12 & 0s & 2m & 4s & 25m & \textbf{10.26}\\
    \bottomrule
    \addlinespace
    \multicolumn{7}{c}{$d=3362$, $N=256$}\\
    \toprule
    Algorithm & $K$ & $NT_{\g}$ & $T_{\f}$ & $T_{\rm model}$ & $T_{\rm alg}$ & $S_{\rm alg}$\\
    \midrule
    Fine & $-$ & $-$ & $-$ & $-$ & 10h 58m & 1\\
    Parareal & 195 & 0s & 2m & 0s & 7h 52m & 1.40\\
    RandNet-Parareal & 12 & 0s & 3m & 20s & 33m & \textbf{19.87}\\
    \bottomrule
    \addlinespace
    \multicolumn{7}{c}{$d=11858$, $N=512$}\\
    \toprule
    Algorithm & $K$ & $NT_{\g}$ & $T_{\f}$ & $T_{\rm model}$ & $T_{\rm alg}$ & $S_{\rm alg}$\\
    \midrule
    Fine* & $-$ & $-$ & $-$ & $-$ & 2d 16h & 1\\
    Parareal & 58 & 1s & 7m & 0s & 6h 59m & 9.23\\
    RandNet-Parareal & 6 & 2s & 8m & 1m & 49m & \textbf{78.44}\\
    \bottomrule
    \addlinespace
    \multicolumn{7}{c}{$d=25538$, $N=512$}\\
    \toprule
    Algorithm & $K$ & $NT_{\g}$ & $T_{\f}$ & $T_{\rm model}$ & $T_{\rm alg}$ & $S_{\rm alg}$\\
    \midrule
    Fine* & $-$ & $-$ & $-$ & $-$ & 7d 16h & 1\\
    Parareal & 27 & 8s & 22m & 0s & 9h 50m & 19.25\\
    RandNet-Parareal & 5 & 9s & 23m & 2m & 1h 57m & \textbf{97.40}\\
    \bottomrule
    \addlinespace
    \multicolumn{7}{c}{$d=53792$, $N=512$}\\
    \toprule
    Algorithm & $K$ & $NT_{\g}$ & $T_{\f}$ & $T_{\rm model}$ & $T_{\rm alg}$ & $S_{\rm alg}$\\
    \midrule
    Fine* & $-$ & $-$ & $-$ & $-$ & 21d 7h & 1\\
    Parareal & 19 & 36s & 1h 0m & 0s & 19h 13m & 26.60\\
    RandNet-Parareal & 4 & 42s & 60m & 4m & 4h 6m & \textbf{124.87}\\
    \bottomrule
    \addlinespace
    \multicolumn{7}{c}{$d=110450$, $N=512$}\\
    \toprule
    Algorithm & $K$ & $NT_{\g}$ & $T_{\f}$ & $T_{\rm model}$ & $T_{\rm alg}$ & $S_{\rm alg}$\\
    \midrule
    Fine* & $-$ & $-$ & $-$ & $-$ & 56d 2h & 1\\
    Parareal & 14 & 3m & 2h 38m & 1s & 1d 14h & 35.84\\
    RandNet-Parareal & 4 & 3m & 2h 37m & 7m & 10h 48m & \textbf{124.52}\\
    \bottomrule
    \addlinespace
    \end{tabular}
    \caption*{Simulation study on the empirical scalability and speed-up of Parareal, nnGParareal (with $m_{\rm nnGP}=20$), and RandNet-Parareal (with $m_{\rm RandNet}=4$ and $M=100$) for the Diffusion-Reaction equation. $T_\f$ and $T_\g$ refer to the runtimes per interval of the fine and coarse solvers, respectively, while $N T_\g$ is the runtime of the coarse solver over $N$ intervals. $T_{\rm model}$ corresponds to the overall time to evaluate $\fhat$, including training and predicting, until convergence at iteration $K$. $T_{\rm alg}$ is the total algorithm runtime, while $S_{\rm alg}$ is the parallel speed-up. ``Fine*'' indicates that the total runtime has been \textit{estimated} extrapolating data from the other algorithms. Missing nnGParareal rows for $d \geq 3362$ are due to convergence failure within a 48-hour time budget.}
    \label{tab:diffreact_speedup}
     \end{table}

\clearpage

\newpage
\section*{NeurIPS Paper Checklist}

\begin{enumerate}

\item {\bf Claims}
    \item[] Question: Do the main claims made in the abstract and introduction accurately reflect the paper's contributions and scope?
    \item[] Answer: \answerYes{} %
    \item[] Justification: All claims, including but not limited to time gains, model training times, numbers of iterations to convergence, speed-ups, and scalability are backed up by empirical results reported in Tables \ref{tab:burgers_summ}, \ref{tab:swe_speedup} and Figure~\ref{fig:diff_react_speedup} in the main text, and Figure \ref{fig:brusselator} and Table \ref{tab:diffreact_speedup} in  Supplementary Material \ref{supp:DiffReactTable}. A comparison between theoretical and empirical results is also provided. These claims are stated in the abstract, introduction, and in the final section.
    \item[] Guidelines:
    \begin{itemize}
        \item The answer NA means that the abstract and introduction do not include the claims made in the paper.
        \item The abstract and/or introduction should clearly state the claims made, including the contributions made in the paper and important assumptions and limitations. A No or NA answer to this question will not be perceived well by the reviewers. 
        \item The claims made should match theoretical and experimental results, and reflect how much the results can be expected to generalize to other settings. 
        \item It is fine to include aspirational goals as motivation as long as it is clear that these goals are not attained by the paper. 
    \end{itemize}

\item {\bf Limitations}
    \item[] Question: Does the paper discuss the limitations of the work performed by the authors?
    \item[] Answer: \answerYes{} %
    \item[] Justification: relevant limitations are discussed in Section~\ref{sec:discussion}, second paragraph. The robustness of the 
   proposed algorithm to several factors, such as the number of neural networks $M$, the number of nearest neighbors $m$ and the sampled random weights $A,\bs{\zeta}$ is introduced at the end of Section \ref{sec:elm}, and investigated in details in Supplementary Material \ref{supp:robustness}. The scaling performance of the algorithm with respect to the number of cores $N$ and model dimensions $d$ is extensively discussed in Section \ref{sec:experiments}.
   Finally, a rescaling of the system is proposed if the data do not meet the condition $||\bs{U}||\leq Q$ in Theorem \ref{prop:approc_bound}.
    \item[] Guidelines:
    \begin{itemize}
        \item The answer NA means that the paper has no limitation while the answer No means that the paper has limitations, but those are not discussed in the paper. 
        \item The authors are encouraged to create a separate "Limitations" section in their paper.
        \item The paper should point out any strong assumptions and how robust the results are to violations of these assumptions (e.g., independence assumptions, noiseless settings, model well-specification, asymptotic approximations only holding locally). The authors should reflect on how these assumptions might be violated in practice and what the implications would be.
        \item The authors should reflect on the scope of the claims made, e.g., if the approach was only tested on a few datasets or with a few runs. In general, empirical results often depend on implicit assumptions, which should be articulated.
        \item The authors should reflect on the factors that influence the performance of the approach. For example, a facial recognition algorithm may perform poorly when image resolution is low or images are taken in low lighting. Or a speech-to-text system might not be used reliably to provide closed captions for online lectures because it fails to handle technical jargon.
        \item The authors should discuss the computational efficiency of the proposed algorithms and how they scale with dataset size.
        \item If applicable, the authors should discuss possible limitations of their approach to address problems of privacy and fairness.
        \item While the authors might fear that complete honesty about limitations might be used by reviewers as grounds for rejection, a worse outcome might be that reviewers discover limitations that aren't acknowledged in the paper. The authors should use their best judgment and recognize that individual actions in favor of transparency play an important role in developing norms that preserve the integrity of the community. Reviewers will be specifically instructed to not penalize honesty concerning limitations.
    \end{itemize}

\item {\bf Theory Assumptions and Proofs}
    \item[] Question: For each theoretical result, does the paper provide the full set of assumptions and a complete (and correct) proof?
    \item[] Answer: \answerYes{} %
    \item[] Justification: Proposition \ref{prop:approc_bound} is clearly stated with all the required assumptions. The proof is not given,  as we cite this result from~\cite{gonon2023approximation}, adjusting their notation to match our, as we clearly mention. The derivation of the computational complexity analysis is provided with all relevant details in Supplementary Material \ref{supp:comp_complx}.
    \item[] Guidelines:
    \begin{itemize}
        \item The answer NA means that the paper does not include theoretical results. 
        \item All the theorems, formulas, and proofs in the paper should be numbered and cross-referenced.
        \item All assumptions should be clearly stated or referenced in the statement of any theorems.
        \item The proofs can either appear in the main paper or the supplemental material, but if they appear in the supplemental material, the authors are encouraged to provide a short proof sketch to provide intuition. 
        \item Inversely, any informal proof provided in the core of the paper should be complemented by formal proofs provided in appendix or supplemental material.
        \item Theorems and Lemmas that the proof relies upon should be properly referenced. 
    \end{itemize}

    \item {\bf Experimental Result Reproducibility}
    \item[] Question: Does the paper fully disclose all the information needed to reproduce the main experimental results of the paper to the extent that it affects the main claims and/or conclusions of the paper (regardless of whether the code and data are provided or not)?
    \item[] Answer: \answerYes{} %
    \item[] Justification: We have taken care of ensuring the reproducibility of all results through a precise use of notation, and by detailing pseudocodes for the algorithms in Supplementary Material \ref{supp:psudocodes}. Additionally, we comprehensively describe the simulation setups both in the main text and in Supplementary Material \ref{supp:sim_setup}.
Moreover, a link to a GitHub repository with a step-by-step Jupyter notebook outlining RandNet-Parareal, and the necessary code to reproduced the results has been provided in Section \ref{sec:intro} in the main text.
    \item[] Guidelines:
    \begin{itemize}
        \item The answer NA means that the paper does not include experiments.
        \item If the paper includes experiments, a No answer to this question will not be perceived well by the reviewers: Making the paper reproducible is important, regardless of whether the code and data are provided or not.
        \item If the contribution is a dataset and/or model, the authors should describe the steps taken to make their results reproducible or verifiable. 
        \item Depending on the contribution, reproducibility can be accomplished in various ways. For example, if the contribution is a novel architecture, describing the architecture fully might suffice, or if the contribution is a specific model and empirical evaluation, it may be necessary to either make it possible for others to replicate the model with the same dataset, or provide access to the model. In general. releasing code and data is often one good way to accomplish this, but reproducibility can also be provided via detailed instructions for how to replicate the results, access to a hosted model (e.g., in the case of a large language model), releasing of a model checkpoint, or other means that are appropriate to the research performed.
        \item While NeurIPS does not require releasing code, the conference does require all submissions to provide some reasonable avenue for reproducibility, which may depend on the nature of the contribution. For example
        \begin{enumerate}
            \item If the contribution is primarily a new algorithm, the paper should make it clear how to reproduce that algorithm.
            \item If the contribution is primarily a new model architecture, the paper should describe the architecture clearly and fully.
            \item If the contribution is a new model (e.g., a large language model), then there should either be a way to access this model for reproducing the results or a way to reproduce the model (e.g., with an open-source dataset or instructions for how to construct the dataset).
            \item We recognize that reproducibility may be tricky in some cases, in which case authors are welcome to describe the particular way they provide for reproducibility. In the case of closed-source models, it may be that access to the model is limited in some way (e.g., to registered users), but it should be possible for other researchers to have some path to reproducing or verifying the results.
        \end{enumerate}
    \end{itemize}

\item {\bf Open access to data and code}
    \item[] Question: Does the paper provide open access to the data and code, with sufficient instructions to faithfully reproduce the main experimental results, as described in supplemental material?
    \item[] Answer: \answerYes{} %
    \item[] Justification: All experimental results are fully reproducible, with code provided via a GitHub repository, with a link shared in Section \ref{sec:intro} in the main text. Each simulation and its corresponding analysis are clearly labeled, and a step-by-step Jupyter notebook is provided to aid the reader in becoming familiar with the API's usage. The repository follows the best practices of the most common ML repositories.
    
    \item[] Guidelines:
    \begin{itemize}
        \item The answer NA means that paper does not include experiments requiring code.
        \item Please see the NeurIPS code and data submission guidelines (\url{https://nips.cc/public/guides/CodeSubmissionPolicy}) for more details.
        \item While we encourage the release of code and data, we understand that this might not be possible, so “No” is an acceptable answer. Papers cannot be rejected simply for not including code, unless this is central to the contribution (e.g., for a new open-source benchmark).
        \item The instructions should contain the exact command and environment needed to run to reproduce the results. See the NeurIPS code and data submission guidelines (\url{https://nips.cc/public/guides/CodeSubmissionPolicy}) for more details.
        \item The authors should provide instructions on data access and preparation, including how to access the raw data, preprocessed data, intermediate data, and generated data, etc.
        \item The authors should provide scripts to reproduce all experimental results for the new proposed method and baselines. If only a subset of experiments are reproducible, they should state which ones are omitted from the script and why.
        \item At submission time, to preserve anonymity, the authors should release anonymized versions (if applicable).
        \item Providing as much information as possible in supplemental material (appended to the paper) is recommended, but including URLs to data and code is permitted.
    \end{itemize}

\item {\bf Experimental Setting/Details}
    \item[] Question: Does the paper specify all the training and test details (e.g., data splits, hyperparameters, how they were chosen, type of optimizer, etc.) necessary to understand the results?
    \item[] Answer: \answerYes{} %
    \item[] Justification: All the experimental setups, training details, and simulation parameters are described in the text, mainly in Sections \ref{sec:elm} and \ref{sec:experiments}. Moreover, they are also summarized in Supplementary Material \ref{supp:sim_setup}.
    \item[] Guidelines:
    \begin{itemize}
        \item The answer NA means that the paper does not include experiments.
        \item The experimental setting should be presented in the core of the paper to a level of detail that is necessary to appreciate the results and make sense of them.
        \item The full details can be provided either with the code, in appendix, or as supplemental material.
    \end{itemize}

\item {\bf Experiment Statistical Significance}
    \item[] Question: Does the paper report error bars suitably and correctly defined or other appropriate information about the statistical significance of the experiments?
    \item[] Answer: \answerNo{} %
    \item[] Justification: %
    We acknowledge that error quantification for the speed-up might be of interest in some situations. However, given the runtime of our experiments, this would be too computationally expensive to obtain. Nevertheless, we reported two robustness studies for two different, smaller systems among the ones considered (Figures \ref{fig:Burgers_perf_across_m} and \ref{fig:DiffReact_perf_across_m}), where the performance of the algorithm is averaged across multiple runs. There, we display the more informative empirical distribution instead of just the error bars.
    \item[] Guidelines:
    \begin{itemize}
        \item The answer NA means that the paper does not include experiments.
        \item The authors should answer "Yes" if the results are accompanied by error bars, confidence intervals, or statistical significance tests, at least for the experiments that support the main claims of the paper.
        \item The factors of variability that the error bars are capturing should be clearly stated (for example, train/test split, initialization, random drawing of some parameter, or overall run with given experimental conditions).
        \item The method for calculating the error bars should be explained (closed form formula, call to a library function, bootstrap, etc.)
        \item The assumptions made should be given (e.g., Normally distributed errors).
        \item It should be clear whether the error bar is the standard deviation or the standard error of the mean.
        \item It is OK to report 1-sigma error bars, but one should state it. The authors should preferably report a 2-sigma error bar than state that they have a 96\%
        \item For asymmetric distributions, the authors should be careful not to show in tables or figures symmetric error bars that would yield results that are out of range (e.g. negative error rates).
        \item If error bars are reported in tables or plots, The authors should explain in the text how they were calculated and reference the corresponding figures or tables in the text.
    \end{itemize}

\item {\bf Experiments Compute Resources}
    \item[] Question: For each experiment, does the paper provide sufficient information on the computer resources (type of compute workers, memory, time of execution) needed to reproduce the experiments?
    \item[] Answer: \answerYes{} %
    \item[] Justification: All results report the execution runtime. Details on the hardware used are provided in Section \ref{sec:intro} in the main text.
    \item[] Guidelines:
    \begin{itemize}
        \item The answer NA means that the paper does not include experiments.
        \item The paper should indicate the type of compute workers CPU or GPU, internal cluster, or cloud provider, including relevant memory and storage.
        \item The paper should provide the amount of compute required for each of the individual experimental runs as well as estimate the total compute. 
        \item The paper should disclose whether the full research project required more compute than the experiments reported in the paper (e.g., preliminary or failed experiments that didn't make it into the paper). 
    \end{itemize}
    
\item {\bf Code Of Ethics}
    \item[] Question: Does the research conducted in the paper conform, in every respect, with the NeurIPS Code of Ethics \url{https://neurips.cc/public/EthicsGuidelines}?
    \item[] Answer: \answerYes{} %
    \item[] Justification: We have reviewed the Code of Ethics and found no particular area of concern regarding our research. 
    \item[] Guidelines:
    \begin{itemize}
        \item The answer NA means that the authors have not reviewed the NeurIPS Code of Ethics.
        \item If the authors answer No, they should explain the special circumstances that require a deviation from the Code of Ethics.
        \item The authors should make sure to preserve anonymity (e.g., if there is a special consideration due to laws or regulations in their jurisdiction).
    \end{itemize}

\item {\bf Broader Impacts}
    \item[] Question: Does the paper discuss both potential positive societal impacts and negative societal impacts of the work performed?
    \item[] Answer: \answerYes{} %
    \item[] Justification: The potential positive impacts are implicitly mentioned in Section~\ref{sec:intro}, and in Section \ref{sec:discussion} (when referring to having chosen systems of real-world significance, with the necessary prerequisites for using the proposed algorithm in practical scenarios). By enabling faster convergence times with minimal overhead, RandNet-Parareal can be applied to a wide range of applications, such as plasma physics simulation, weather forecasting (both mentioned in the introduction), leading to positive societal impacts. 
    \item[] Guidelines:
    \begin{itemize}
        \item The answer NA means that there is no societal impact of the work performed.
        \item If the authors answer NA or No, they should explain why their work has no societal impact or why the paper does not address societal impact.
        \item Examples of negative societal impacts include potential malicious or unintended uses (e.g., disinformation, generating fake profiles, surveillance), fairness considerations (e.g., deployment of technologies that could make decisions that unfairly impact specific groups), privacy considerations, and security considerations.
        \item The conference expects that many papers will be foundational research and not tied to particular applications, let alone deployments. However, if there is a direct path to any negative applications, the authors should point it out. For example, it is legitimate to point out that an improvement in the quality of generative models could be used to generate deepfakes for disinformation. On the other hand, it is not needed to point out that a generic algorithm for optimizing neural networks could enable people to train models that generate Deepfakes faster.
        \item The authors should consider possible harms that could arise when the technology is being used as intended and functioning correctly, harms that could arise when the technology is being used as intended but gives incorrect results, and harms following from (intentional or unintentional) misuse of the technology.
        \item If there are negative societal impacts, the authors could also discuss possible mitigation strategies (e.g., gated release of models, providing defenses in addition to attacks, mechanisms for monitoring misuse, mechanisms to monitor how a system learns from feedback over time, improving the efficiency and accessibility of ML).
    \end{itemize}
    
\item {\bf Safeguards}
    \item[] Question: Does the paper describe safeguards that have been put in place for responsible release of data or models that have a high risk for misuse (e.g., pretrained language models, image generators, or scraped datasets)?
    \item[] Answer: \answerNA{} %
    \item[] Justification: We relied on publicly available models, simulating the relevant data as described in the main text and in Supplementary Material. %
    \item[] Guidelines:
    \begin{itemize}
        \item The answer NA means that the paper poses no such risks.
        \item Released models that have a high risk for misuse or dual-use should be released with necessary safeguards to allow for controlled use of the model, for example by requiring that users adhere to usage guidelines or restrictions to access the model or implementing safety filters. 
        \item Datasets that have been scraped from the Internet could pose safety risks. The authors should describe how they avoided releasing unsafe images.
        \item We recognize that providing effective safeguards is challenging, and many papers do not require this, but we encourage authors to take this into account and make a best faith effort.
    \end{itemize}

\item {\bf Licenses for existing assets}
    \item[] Question: Are the creators or original owners of assets (e.g., code, data, models), used in the paper, properly credited and are the license and terms of use explicitly mentioned and properly respected?
    \item[] Answer: \answerYes{} %
    \item[] Justification: All creators have been properly credited both in terms of published scientific papers, and publicly available code and libraries (e.g. for some specific Python libraries).
    \item[] Guidelines:
    \begin{itemize}
        \item The answer NA means that the paper does not use existing assets.
        \item The authors should cite the original paper that produced the code package or dataset.
        \item The authors should state which version of the asset is used and, if possible, include a URL.
        \item The name of the license (e.g., CC-BY 4.0) should be included for each asset.
        \item For scraped data from a particular source (e.g., website), the copyright and terms of service of that source should be provided.
        \item If assets are released, the license, copyright information, and terms of use in the package should be provided. For popular datasets, \url{paperswithcode.com/datasets} has curated licenses for some datasets. Their licensing guide can help determine the license of a dataset.
        \item For existing datasets that are re-packaged, both the original license and the license of the derived asset (if it has changed) should be provided.
        \item If this information is not available online, the authors are encouraged to reach out to the asset's creators.
    \end{itemize}

\item {\bf New Assets}
    \item[] Question: Are new assets introduced in the paper well documented and is the documentation provided alongside the assets?
    \item[] Answer: \answerYes{} %
    \item[] Justification: The code, simulations and associated analyses are publicly released with permissive licence.
    \item[] Guidelines:
    \begin{itemize}
        \item The answer NA means that the paper does not release new assets.
        \item Researchers should communicate the details of the dataset/code/model as part of their submissions via structured templates. This includes details about training, license, limitations, etc. 
        \item The paper should discuss whether and how consent was obtained from people whose asset is used.
        \item At submission time, remember to anonymize your assets (if applicable). You can either create an anonymized URL or include an anonymized zip file.
    \end{itemize}

\item {\bf Crowdsourcing and Research with Human Subjects}
    \item[] Question: For crowdsourcing experiments and research with human subjects, does the paper include the full text of instructions given to participants and screenshots, if applicable, as well as details about compensation (if any)? 
    \item[] Answer: \answerNA{} %
    \item[] Justification: The paper involves neither crowdsourcing nor research with human subjects.
    \item[] Guidelines:
    \begin{itemize}
        \item The answer NA means that the paper does not involve crowdsourcing nor research with human subjects.
        \item Including this information in the supplemental material is fine, but if the main contribution of the paper involves human subjects, then as much detail as possible should be included in the main paper. 
        \item According to the NeurIPS Code of Ethics, workers involved in data collection, curation, or other labor should be paid at least the minimum wage in the country of the data collector. 
    \end{itemize}

\item {\bf Institutional Review Board (IRB) Approvals or Equivalent for Research with Human Subjects}
    \item[] Question: Does the paper describe potential risks incurred by study participants, whether such risks were disclosed to the subjects, and whether Institutional Review Board (IRB) approvals (or an equivalent approval/review based on the requirements of your country or institution) were obtained?
    \item[] Answer: \answerNA{} %
    \item[] Justification: The paper involves neither crowdsourcing nor research with human subjects.
    \item[] Guidelines:
    \begin{itemize}
        \item The answer NA means that the paper does not involve crowdsourcing nor research with human subjects.
        \item Depending on the country in which research is conducted, IRB approval (or equivalent) may be required for any human subjects research. If you obtained IRB approval, you should clearly state this in the paper. 
        \item We recognize that the procedures for this may vary significantly between institutions and locations, and we expect authors to adhere to the NeurIPS Code of Ethics and the guidelines for their institution. 
        \item For initial submissions, do not include any information that would break anonymity (if applicable), such as the institution conducting the review.
    \end{itemize}

\end{enumerate}

\end{document}

%% file: commands.tex
\newcommand{\R}{\mathbb{R}}

\usepackage{mathrsfs}
\newcommand{\f}{\mathscr{F}}
\newcommand{\g}{\mathscr{G}}

\newcommand{\fhat}{\widehat{f}}

\mathchardef\mhyphen="2D 

\newcommand{\bs}{\boldsymbol}

\newtheorem{proposition}{Proposition}